\documentclass[aps,superscriptaddress,nofootinbib,notitlepage,twocolumn]{revtex4-1}

\usepackage[latin1]{inputenc}
\usepackage{graphicx}
\newcommand\aap{{\rm A\&A}}
\newcommand\mnras{{\rm MNRAS}}
\usepackage{float}
\usepackage{latexsym}
\usepackage{graphicx}
\usepackage{amssymb}
\usepackage{amsmath}
\usepackage{amsfonts}
\usepackage{hyperref}
\usepackage{cool}
\usepackage{dcolumn}
\usepackage{textcomp}
\usepackage{color}
\usepackage{xfrac}
\usepackage{slashed}
\usepackage{multirow}
\usepackage{comment}

\def\be{\begin{equation}}
\def\ee{\end{equation}}
\def\figs/B{B}
\def\bea{\begin{eqnarray}}
\def\eea{\end{eqnarray}}
\def\bg{\begin{eqnarray}}
\def\nd{\end{eqnarray}}

\def\cos{{\rm cos}}

\newcommand\jcap{{\rm JCAP}}

\definecolor{orange}{rgb}{1,0.5,0}

\def\be{\begin{equation}}
\def\ee{\end{equation}}

\begin{document}

\title{A Dark Matter Trigger for Early Dark Energy Coincidence}

\begin{abstract}

Early dark energy (EDE), whose cosmological role is localized in time around the epoch of matter-radiation equality in order to resolve the Hubble tension, introduces a new coincidence problem: why should the EDE dynamics occur near equality if EDE is decoupled from both matter and radiation?  The resolution of this problem may lie in an {\it early dark sector} (EDS), wherein the dark matter mass is dependent on the EDE scalar field. Concretely, we consider a Planck-suppressed coupling of EDE to dark matter, as would naturally arise from breaking of the global $U(1)$ shift symmetry of the former by quantum gravity effects. With a sufficiently flat potential, the rise to dominance of dark matter at matter-radiation equality itself triggers the rolling and subsequent decay of the EDE. We show that this {\it trigger} EDS (tEDS) model can naturally resolve the EDE coincidence problem at the background level without any fine tuning of the coupling to dark matter or of the initial conditions.  When fitting to current cosmological data, including that from the local distance ladder and the low-redshift amplitude of fluctuations, the tEDS maximum-likelihood model performs comparably to EDE for resolving the Hubble tension, achieving $H_0 =71.2$ km/s/Mpc.  However, fitting the \emph{Planck} cosmic microwave background data requires a specific range of initial field positions to balance the scalar field fluctuations that drive acoustic oscillations, providing testable differences with other EDE models.  

\end{abstract}

\author{Meng-Xiang Lin}
\affiliation{Kavli Institute for Cosmological Physics and Enrico Fermi Institute, The University of Chicago, Chicago, IL 60637, USA}
\affiliation{Center for Particle Cosmology, Department of Physics and Astronomy, University of Pennsylvania, Philadelphia, PA 19104, USA}

\author{Evan McDonough}
\affiliation{Kavli Institute for Cosmological Physics and Enrico Fermi Institute, The University of Chicago, Chicago, IL 60637, USA}
\affiliation{Department of Physics, University of Winnipeg, Winnipeg, MB R3B 2E9 Canada}

\author{J.~Colin Hill}
\affiliation{Department of Physics, Columbia University, New York, NY, USA 10027}

\author{Wayne Hu}
\affiliation{Kavli Institute for Cosmological Physics and Enrico Fermi Institute, The University of Chicago, Chicago, IL 60637, USA}


\maketitle

\section{A New Coincidence Problem}\label{sec:intro}

The discrepancy between the inference  of the Hubble constant from the {\it Planck} cosmic microwave background (CMB) data  \cite{Aghanim:2018eyx} within the $\Lambda$CDM model and that from the SH0ES cosmic distance ladder calibrated with Cepheid variables stands at 5$\sigma$ \cite{Riess:2021jrx}.  More generally, many direct probes of $H_0$ also disagree with the CMB (and large-scale structure constraints) at varying levels of quoted significance \cite{Verde:2019ivm,Kamionkowski:2022pkx} (see, however,~\cite{Freedman:2019jwv,Freedman:2021ahq}). This ``Hubble tension'' has spurred on an intense effort in the cosmology community, aiming to explain the origin of this discrepancy via new physics in the cosmological model. 

A prominent approach to reconcile these measurements is to modify the cosmological model in the pre-recombination universe, thereby changing the CMB inference of $H_0$.  In particular, one may consider a model which reduces the physical size of the sound horizon at last scattering, $r_s$, relative to that found in the concordance $\Lambda$CDM model; since the CMB data constrain the {\it angular} size of the sound horizon, $\theta_s$, such a change then raises the $H_0$ value inferred from the CMB. A paradigmatic example within this class of models is early dark energy (EDE), originally proposed by \cite{Karwal:2016vyq,Poulin:2018cxd} and followed by many other realizations, e.g., Refs.~\cite{Agrawal:2019lmo,Lin:2019qug,Smith:2019ihp,Alexander:2019rsc,Alexander:2022own} (see \cite{Kamionkowski:2022pkx} for a recent review).

The size of the sound horizon $r_s$ and that of the CMB damping scale are most sensitive, in different ways, to the decade of redshift that precedes last scattering \cite{Knox:2019rjx}. Meanwhile, measurements of the post-recombination universe are in excellent agreement with $\Lambda$CDM, with relatively little room for modification. This suggests that the epoch of matter-radiation equality occupies a privileged position in the hierarchy of time-scales in the new model in order to mimic the success of $\Lambda$CDM \cite{Karwal:2016vyq}. 

This presents a new cosmological mystery, which we refer to as the {\it Early Dark Energy Coincidence Problem}: 
\begin{quote}  
{\it
The resolution of the Hubble tension by a dark energy-like component crucially relies on a coincidence of unknown origin, namely, that its epoch of influence corresponds to the time of matter-radiation equality.
} 
\end{quote}
This may easily be observed in all successful EDE-like models (see, e.g.,~\cite{Poulin:2018cxd,Lin:2019qug,Smith:2019ihp}). In these models, the new dark-energy-like component contributes a fraction of the energy density of the universe, $f_{\rm EDE}$, that is sharply peaked at a redshift $z_c$. If the model is to address the Hubble tension, then $z_c$ must be $\approx z_{\rm eq}$ which would appear to be a coincidence lacking a dynamical explanation.

The EDE coincidence problem has been studied in a small number of works with a limited amount of success. Concretely, Ref.~\cite{Sabla:2021nfy} obtained a no-go with assisted quintessence; Refs.~\cite{Sakstein:2019fmf,CarrilloGonzalez:2020oac} proposed neutrino-assisted EDE, which has the potential to resolve the coincidence problem, but has yet to be tested against observables and may violate existing constraints on neutrino masses; and Ref.~\cite{Karwal:2021vpk} study the interaction between EDE and dark matter in a specific scenario that fails to address the coincidence problem, as described in further detail below.

The early dark sector (EDS) scenario \cite{McDonough:2021pdg} encompasses 
a general interaction between EDE and dark matter, in the form of an EDE-dependent mass for the dark matter particle.  It has been studied in a specific incarnation as a mechanism to reduce the disagreement in the EDE scenario between Hubble-tension-resolving models that fit the CMB data and observations of large-scale structure \cite{Hill:2020osr,Ivanov:2020ril,DAmico:2020ods,Jedamzik:2020zmd,Lin:2021sfs} (see also \cite{Smith:2020rxx,Murgia:2020ryi,Simon:2022adh,Herold:2021ksg}). 
One might also wonder, in a more general context, whether EDS can address the EDE coincidence problem.

In this work we answer the EDE coincidence question within the EDS framework in the affirmative: we demonstrate that an EDE-dependent perturbative correction to the dark matter mass, itself a generic expectation of effective field theory, can act as a {\it trigger} for the decay of the EDE that naturally occurs at matter-radiation equality.  We provide three detailed conditions that any coincidence-free model must satisfy, and demonstrate that our trigger EDS (tEDS) model satisfies them all.

The structure of this paper is as follows: 
In Sec.~\ref{sec:bg} we enumerate the model requirements on the background cosmological evolution for a successful resolution of the EDE coincidence problem, and propose the tEDS model which satisfies these requirements. 
In Sec.~\ref{sec:data} we confront the model with data from \emph{Planck} 2018 CMB, BOSS BAO, Pantheon SNIa, and the SH0ES cosmic distance ladder measurement of $H_0$, and find that the tEDS model performs comparably well to EDE. 
We investigate this further in Sec.~\ref{sec:pert}, where we perform a detailed analysis of CMB physics in the tEDS model, and show why a specific range of initial field values is preferred in spite of the insensitivity of the trigger mechanism. 
In Sec.~\ref{sec:LSS} we show that tEDS also has the (partial) ability to reduce $S_8$ compared to that found in EDE, from which it benefits once Dark Energy Survey data is included.  
We discuss results and directions for future work in Sec.~\ref{sec:discussion}.

\section{Resolving the Coincidence Problem}\label{sec:bg}

The EDE coincidence problem finds a natural solution in the context of the EDS scenario. EDS was originally proposed in \cite{McDonough:2021pdg} as a mechanism to resolve the Hubble tension without exacerbating the currently moderate tension between CMB and large-scale structure inferences of $S_8$ within $\Lambda$CDM. 
The EDS scenario posits an interaction between dark matter (DM) and the EDE field, such that the mass of the DM particle is dependent on the EDE scalar. 

While past work \cite{McDonough:2021pdg} considered the limit in which the dominant effect of the EDE-DM interaction is in the DM background and perturbations, with only a small backreaction on the evolution of the EDE scalar, in the present work we are interested in the opposite regime: a small change in the DM mass that has a dramatic effect on the background evolution of the EDE scalar. In this regime, the dark matter can act as a {\it trigger} for the decay of the EDE so that the scalar is naturally released from Hubble drag near matter-radiation equality.

EDS models are specified by a potential $V(\phi)$ and a dark matter mass $m_{\rm DM}(\phi)$. The background evolution for the DM energy density $\rho_{\rm DM}$ is given by 
\begin{equation}\label{eq:rhoDMwithint}
    \dot{\rho}_{\rm DM} + 3 a H \rho_{\rm DM} = \dot{\phi}\frac{d \ln m_{\rm DM}}{d \phi}\rho_{\rm DM} ,
\end{equation}
and for the EDE scalar $\phi$ by
\begin{equation}\label{eq:KG}
    \ddot{\phi} + 2 a H \dot{\phi} +  a^2 V'_{\rm eff} =0,
\end{equation}
where overdots are derivatives with respect to conformal time and the Hubble parameter is defined with respect to the coordinate time, $H \equiv d\ln a/dt$. 
Here $V_{\rm eff}$ is an effective potential that includes both the ``bare'' potential $V(\phi)$ and the interaction with DM, defined by
\begin{equation}
    V'_{\rm eff} = V' +  \rho_{\rm DM} \frac{d\ln m_{\rm DM}}{d \phi}
\label{eq:phiEOM}
\end{equation}
where $^\prime$ is the derivative with respect to $\phi$.
The effective potential dictates the transition in the EDE evolution from dark-energy-like (i.e., a cosmological constant) to decaying. The timing of this transition can be approximated as the epoch when
\begin{equation}
    \frac{V'_{\rm eff}}{H^2 \phi} \sim {\cal O}(1) \,\,\, {\rm (onset\, of \, rolling)}
\end{equation}
is first satisfied, corresponding to the release of the field from Hubble drag (cf. Eq.~\eqref{eq:KG} above).   If $|V'_{\rm eff}| \gg |V'|$ at this time, the release and subsequent decay of the EDE can be considered to be ``triggered'' by the DM coupling.  

Following past work, we parametrize the evolution of the EDE in terms of its fractional contribution  
\begin{equation}
    f_{\rm EDE}(z)  \equiv \frac{\rho_{\rm EDE}(z)}{ \rho_{\rm tot}(z)} \;\; , \;\; \rho_{\rm EDE} \equiv \frac{1}{2}a^{-2}\dot{\phi^2} + V(\phi)
\end{equation}
to the total energy density $\rho_{\rm tot}$.
This is maximal at a critical redshift $z_c$, and we will frequently refer to $f_{\rm EDE}(z_c)$ as simply $f_{\rm EDE}$.

\subsection{Model Requirements}
\label{sec:modelreqs}
We make the following demands on a model for it to be deemed a resolution of the Hubble tension free from the EDE coincidence problem:
\begin{enumerate}
    \item Resolve the Hubble tension via EDE-like dynamics, with $f_{\rm EDE} \sim 0.1$ and $z_c\sim10^{3.5}$ (approximately matter-radiation equality),
    and a release from Hubble drag
    \begin{equation}
    \Big|\frac{d\ln \phi}{d\ln a}\Big| \sim \Big|\frac{V'_{\rm eff}}{ \phi \, H^2}\Big| \sim 1
        \label{eq:release}
    \end{equation}
    just prior to this epoch. 
    
    \item DM-triggered decay: The release from Hubble drag is triggered by the coupling  rather than the bare potential $V(\phi)$, i.e., $|V_{\rm eff}^{\prime}|\gg|V^{\prime}|$, implying 
    \begin{equation}
    \frac{\rho_{\rm DM} }{H^2 \phi} \Big| \frac{d\ln m_{\rm DM}}{ d\phi} \Big| \sim 1
    \label{eq:generictrigger}
    \end{equation}
    at release.
    
    \item No fine-tuning of initial conditions: The mechanism of DM-induced release from Hubble drag at the critical redshift $z_c$ is independent of the initial value of $\phi$. 
\end{enumerate}
These requirements significantly narrow the possibilities for both the mass $m_{\rm DM}(\phi)$ and the potential $V(\phi)$. 

\subsection{Previous Models}

Previous models are unable to satisfy the conditions for a successful trigger mechanism.  For example, Ref.~\cite{Karwal:2021vpk} proposed a monomial potential $V(\phi)=V_0 (\phi/M_{\rm pl})^n$ with $n>2$ and an exponential coupling $m_{\rm DM}(\phi)=m_0 e^{c \phi/M_{\rm pl}}$ as an EDE-like model with the possibility to resolve  the coincidence problem.
The effective potential in this case has slope given by $V'_{\rm eff} = V' + c \rho_{\rm DM}/M_{\rm pl}$. 
Let us confront this with the above conditions:\\

Condition 1, namely, that $z_c \sim z_{\rm eq}$, sets the Hubble parameter at that time as $3 M_{\rm pl}^2 H^2(z_c) \sim \rho_{\rm DM}$. This can be related to $V$ via $f_{\rm EDE}\sim V/ (3 H^2 M_{\rm pl}^2)$ as $ V \sim f_{\rm EDE} \rho_{\rm DM}$, and, using $V=V_0 (\phi/M_{\rm pl})^n$, this fixes the slope $V'\sim n f_{\rm EDE }\rho_{\rm DM}/\phi$.\\

Condition 2 implies that  $c \rho_{\rm DM}/M_{\rm pl} \gg V'$.  Combined with $V'$ from Condition 1, this implies $c \gg n (M_{\rm pl}/\phi)f_{\rm EDE}$.   At release, 
\begin{equation}
    \frac{\rho_{\rm DM}c}{H^2 \phi M_{\rm pl}} \sim 1.
\label{eq:triggerfieldvalue}
\end{equation}

Condition 3 is violated since the trigger mechanism works only around a specific initial value of $\phi$ given by Eq.~(\ref{eq:triggerfieldvalue}).

In addition, even only requiring Conditions 1 and 2 causes problematic phenomenology for satisfying observations.  First, they require a large fractional change in the DM mass between the initial epoch and today:
$|\Delta \ln m | \sim |c \phi/M_{\rm pl} |\gg n f_{\rm EDE}$.
Furthermore, since $\rho_{\rm DM} \sim 3 M_{\rm  pl}^2 H_{\rm eq}^2$, the specific initial value required is $\phi \sim 3 c M_{\rm pl}$ and
even marginally satisfying the trigger condition implies a large coupling $3 c^2 \gtrsim n f_{\rm EDE}$, which is also not observationally favored in this model, in part due to the ``fifth force'' in the dark sector and the enhanced growth that it mediates \cite{McDonough:2021pdg} (see \S \ref{sec:perteq} below).

Similar considerations for the no-go for trigger solutions also apply to the original EDE axion-like potentials, where
\begin{equation}
    V(\theta=\phi/f) = V_0 (1 - \cos \,\theta)^3 \,,
\label{eq:EDEpotential}
\end{equation}
as studied in Ref.~\cite{McDonough:2021pdg} with the same exponential EDE-DM coupling as in Ref.~\cite{Karwal:2021vpk}.

We therefore conclude that with EDE axion-like potentials and an exponential coupling to the DM mass there is no viable resolution to the EDE coincidence problem.

\subsection{Trigger Model}

The requirement that the EDE field is triggered by the DM for generic initial field values from Eq.~(\ref{eq:generictrigger}) suggests that we need a coupling where
$d\ln m_{\rm DM} /d\phi \propto \phi$ with a sufficiently flat bare potential, unlike previous models.  In addition, this form of the coupling makes the fifth force enhancement of growth in the DM sector vanish as $\phi\rightarrow 0$, and can thus remove the consequent late-time enhancement of large-scale structure found in \cite{McDonough:2021pdg} (see \S\ref{sec:perteq}).
 
As a simple, theoretically well-motivated coupling that satisfies this requirement, we consider an EDE-dependence of the dark matter mass given by
\begin{equation}
\label{eq:mDM_quad}
    m_{\rm DM}(\phi)  = m_0 \left( 1 + g\frac{\phi^2}{M_{\rm pl}^2} \right) .
\end{equation}
The interaction with $\phi$ is naturally Planck-suppressed, and the coupling constant $g$ is expected to be an ${\cal O}(1)$ number based on standard effective field theory arguments. This coupling is consistent with the symmetries of the low-energy effective field theory (wherein $\phi$ is typically associated with a pseudo-scalar), and is also a natural expectation of string theory.  For example, the non-perturbative effects in string theory (presumably responsible for the EDE potential) are themselves in general moduli-dependent (see, e.g., Refs.~\cite{Baumann:2006th,Ruehle:2017one}), which could generate couplings of the form given above.  

For the potential, one may again take guidance from effective field theory considerations. The EDE scalar field is naturally identified as an axion-like particle, e.g., the phase of a complex scalar field in a field theory describing the spontaneous breaking of a global $U(1)$ symmetry. In this case the requisite extremely small mass of $\phi$ can be justified on the basis of a continuous shift symmetry that is broken to a discrete shift symmetry non-perturbatively, with the potential protected from perturbative corrections. The potential in this case is periodic, e.g., $V(\theta = \phi/f) = V_0 \, \cos \, \theta$.

In the present context, the requirement that the minimum of the potential is locally $V \propto \phi^{2n}$ with an integer $n\geq 2$ in order for $f_{\rm EDE}$ to decay sufficiently quickly~(e.g.,~\cite{Poulin:2018cxd,Agrawal:2019lmo}), whilst $V'_{\rm eff} \gg V'$ at early times to trigger off the DM, suggests instead a potential of the {\it monodromy} type, namely, a potential that breaks the shift symmetry entirely. In this scenario, the potential is generically {\it flattened} at large field values \cite{Dong:2010in}. For example, the original models of axion monodromy inflation \cite{McAllister:2008hb,Silverstein:2008sg} (see also Ref.~\cite{McAllister:2014mpa}) may be parametrized as
\begin{equation}
\label{eq:Vtheta1}
    V(\theta= \phi/f) =  V_0 (1+ \theta^a)^{\frac{1}{b}},
\end{equation}
where $V_0$ is a normalization.
This potential is characterized by a minimum which is locally $V \sim \theta^a$, and a {\it flattening} of the potential at large field values, $V\sim \theta^{a/b}$ for $a>0$ and $b>1$.  This class of potential has well-developed interesting associated phenomenology, such as the production of oscillons \cite{Amin:2011hj,Lozanov:2016hid,Lozanov:2017hjm}.

More recent axion monodromy models (e.g.,~Ref.~\cite{Blumenhagen:2015qda}) have extended the scenario to include {\it plateau}-like potentials, characterized by an ever more dramatic flattening at large field values. This class of monodromy potential naturally satisfies $V'_{\rm eff} \gg V'$ for arbitrary initial conditions, and thus is a natural candidate for a coincidence-free EDS model. 
To make contact with the EDE literature and enforce a flat plateau, we remap the parameters of Eq.~(\ref{eq:Vtheta1}) as $a=-2 np$ and $b=-p$ with positive integer values of $n$ and $p$. 
This rewriting of the potential makes manifest a plateau $V\rightarrow V_0$ at large field values ($|\theta| \gg 1$), a minimum $V \propto \theta^{2 n}$ at small field values ($|\theta| \ll 1$), and a transition region at $|\theta| \sim 1$. 
Motivated by the phenomenologically successful EDE potential in Eq.~\eqref{eq:EDEpotential}, which behaves as $V\sim\theta^6$ near its minimum, we fix $n=3$ hereafter so that $p$ alone controls the sharpness of the transition.  Our bare potential is therefore parameterized as 
\begin{equation}
\label{eq:V}
    V(\theta=\phi/f) = V_0 \frac{\theta^{6}}{\left(1+\theta^{6p}\right) ^{\frac{1}{p}}}.
\end{equation}

This trigger EDS  model admits a simple interpretation.   For $g \phi^2/M_{\rm pl}^2 \ll 1$, corresponding to a small fractional change in the DM mass, one may approximate $\ln m_{\rm DM} \sim \ln m_0 + g \phi^2/M_{\rm pl}^2$. The effective potential may then be approximated as 
\begin{equation} 
\label{eq:Veff}
V_{\rm eff } \approx V(\phi)+g \frac{\phi^2}{M_{\rm pl}^2} \rho_{\rm DM},
\end{equation}
in effect a correction to the mass of $\phi$,
\begin{equation}
   { m^2 _\phi} _{\rm eff} \approx V '' + {2}g \frac{\rho_{\rm DM}}{M_{\rm pl}^2} .
\end{equation}
Note we use the exact expressions for $V(\phi)$ and $m_{\rm DM}(\phi)$, Eqs.~\eqref{eq:V} and \eqref{eq:mDM_quad}, in all numerical calculations.

\begin{figure}
    \centering
    \includegraphics[width=0.99\columnwidth]{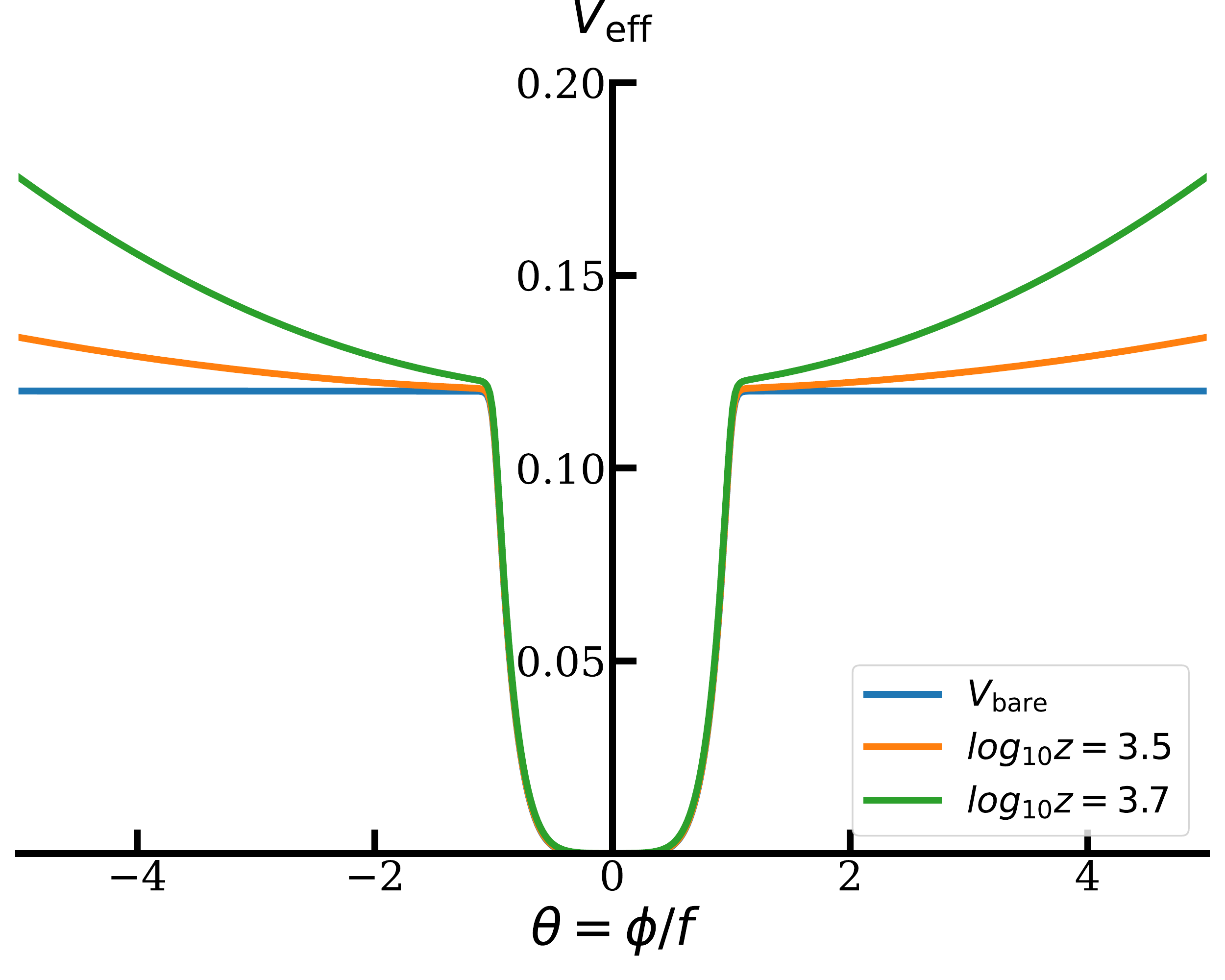}
    \caption{The tEDS bare potential (Eq.~\eqref{eq:V}) versus effective potential (Eq.~\eqref{eq:Veff}) at different redshifts near matter-radiation equality (as labeled). 
    Here we take $p=8$, $f=0.05 M_{\rm pl}$, $V_0=0.12\,\,{\rm eV}^4$, $g=0.68$, and $\Omega_ch^2=0.1280$ as an example.  
    The rolling of the EDE scalar field is triggered by the coupling to dark matter near $z_{\rm eq}$ for any initial field position on the plateau of the bare potential. 
    }
    \label{fig:Veff}
\end{figure}

The effective potential in Eq.~\ref{eq:Veff} is shown in Fig.~\ref{fig:Veff} for $p=8$.
One may immediately understand the dynamics of the model: if the field is initially on the plateau of $V(\phi)$, then $V_{\rm eff}\propto \phi^2$ and the effective mass of $\phi$ is predominantly due to the DM contribution. The field begins to roll when $ { m^2 _\phi} _{\rm eff} \sim  H^2$, which occurs when $H^2 \sim g\rho_{\rm DM}/ M_{\rm pl}^2$.
For $g \sim {\cal O}(1)$, this is satisfied around matter-radiation equality for any initial $\phi$. This construction naturally satisfies all three trigger conditions:  $ V'_{\rm eff}/(\phi H^2)\sim m^2_{\phi_{\rm eff}}/H^2$ is independent of $\phi$ and hence the onset of rolling is independent of initial conditions, satisfying condition 3; $V(\phi)$ for $\phi\gg f$ is flat by design and hence $V'$ is automatically small, satisfying condition 2; and $f_{\rm EDE} \sim V_0/( H^2 M_{\rm pl}^2)$ may be adjusted by setting $V_0$, while $z_c$ is determined by $g$, thus satisfying condition 1.  
Notice that the conditions themselves do not set requirements on the field scale $f$.   This flexibility allows for the ability to adjust the variation in the dark matter mass between $z_c \sim z_{\rm eq}$ and today, $m_{\rm DM}(\theta=1)-m_0 = m_0 g f^2/M^2_{\rm pl}$, and in particular to reduce it while increasing the frequency of field oscillations around the potential minimum at $z\lesssim z_c$.

\begin{figure}
    \centering
    \includegraphics[width=0.99\columnwidth]{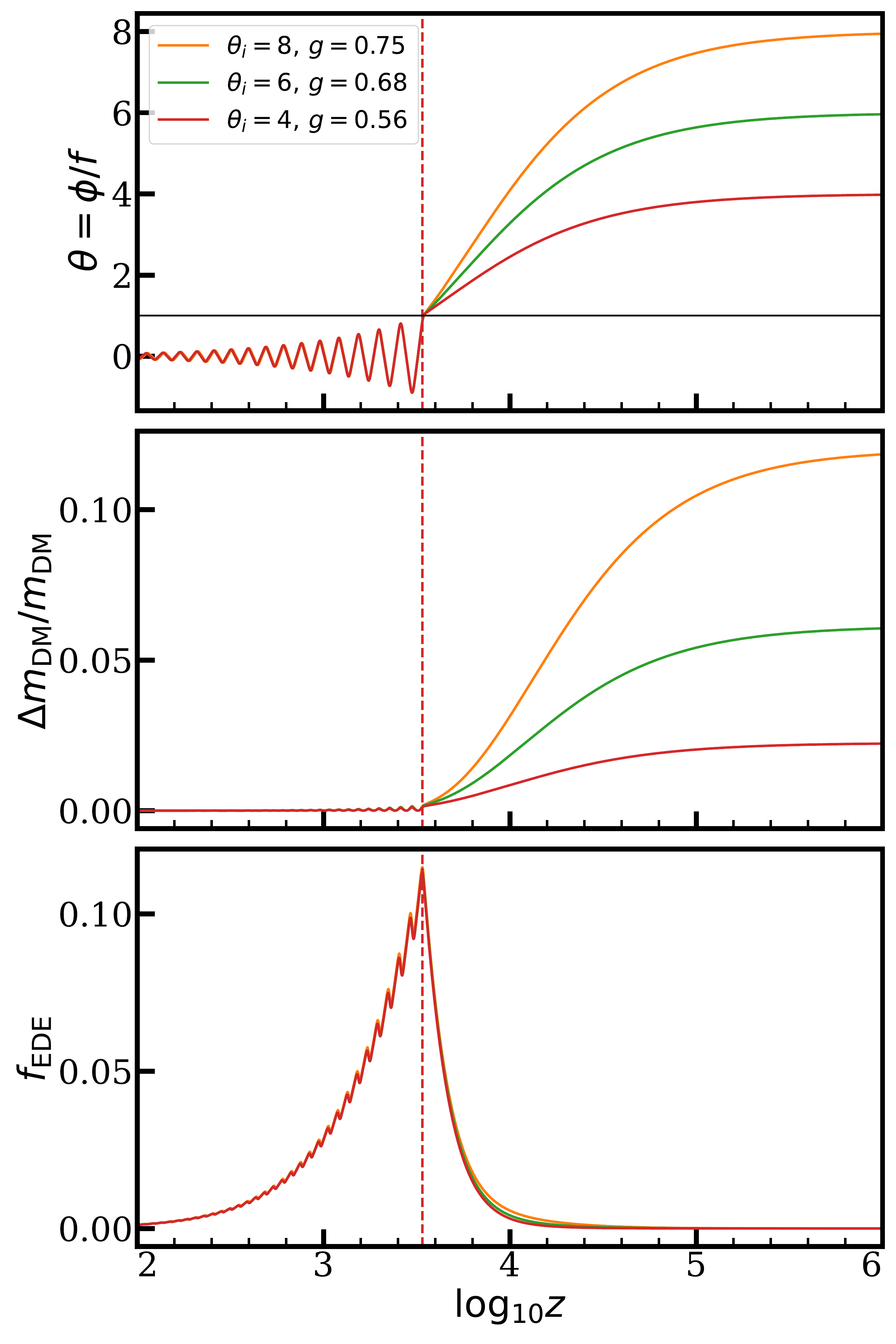}
    \caption{
    The time evolution of the EDE scalar field (top), dark matter mass (middle), and EDE fractional energy density $f_{\rm EDE}(z)$ (bottom) for the tEDS model ($p=8$) with various initial field values $\theta_i$.
    The vertical dashed line indicates the peak epoch $z_c$ and the black horizontal line on the top panel indicates the edge of the potential plateau $\theta=1$.
    For a wide range of initial field positions $\theta_i$, the EDS order-unity coupling $g$ is adjusted fractionally as labeled to achieve a similar $f_{\rm EDE}$ evolution.
    The other parameters are fixed to be the same as in Fig.~\ref{fig:Veff}. 
    }
    \label{fig:theta}
\end{figure}

We confirm these dynamics in Fig.~\ref{fig:theta}, where we numerically solve for the evolution of the background fields in the EDS model defined by Eqs.~\eqref{eq:mDM_quad} and \eqref{eq:V}. In the top panel of Fig.~\ref{fig:theta} we show the evolution of the EDE scalar $\phi$ for varying initial field positions $\theta_i$, for the fixed $p=8$, $f$, and $V_0$ values of Fig.~\ref{fig:Veff}.  The EDE fractional energy density $f_{\rm EDE}(z)$ is shown in the bottom panel of Fig.~\ref{fig:theta}. 

We can see that the Hubble tension target ($f_{\rm EDE}\sim 0.1,\; z_c\sim 10^{3.5}$) can be realized by a wide range of initial conditions. The time evolution of $f_{\rm EDE}(z)$ is not sensitive to the initial field position $\theta_i$ once $z_c$ is fixed by adjusting the coupling constant $g$ by a fractional amount around its typical order-unity magnitude.   In addition, the evolution of the DM mass, especially in the observationally relevant regime $z< z_{\rm eq}\sim z_c$, remains very small (see the middle panel of Fig.~\ref{fig:theta}).  Moreover, since $\phi\rightarrow 0$ the additional fifth force acting on the DM at late times is suppressed as well.

We have also verified that this trigger mechanism and its consequences for the dark matter apply to any sufficiently large $p$ or equivalently any sufficiently flat plateau in $V(\phi)$.

Therefore, we conclude that we can satisfy all the requirements for solving the coincidence problem for the EDE background evolution with the model defined by Eqs.~\eqref{eq:mDM_quad} and \eqref{eq:V}, which we shall refer to as the trigger EDS or tEDS model.

\subsection{Perturbations}
\label{sec:perteq}

A successful resolution to the Hubble tension with the tEDS model must not only reproduce the background energy density features of EDE, but also produce perturbations which fit CMB and large-scale structure data.

Following Ref.~\cite{McDonough:2021pdg}, the equations of motion for the perturbations in synchronous gauge are modified to account for a general coupling as:
\begin{eqnarray}
    \ddot{\delta \phi}&&+ 2 a H \dot{\delta \phi}
    +\left(k^{2} + a^{2}\frac{d^{2}V}{d\phi^{2}}\right)\delta \phi
    +\frac{1}{2}\dot{h}\dot{\phi}= \nonumber \\
    &&-a^2 \left[\frac{d\ln m}{d\phi}\rho_{\rm DM}\delta 
    +\frac{d^{2} \ln m}{d\phi^{2}}\delta\phi\rho_{\rm DM}\right],
\label{eq:deltaphiddot}
\end{eqnarray} 
\begin{eqnarray}\label{denscont}
    \dot{\delta}  + \theta + \frac{\dot{h}}{2}=\frac{d \ln m}{d \phi}\dot{\delta \phi}+ \frac{d^2 \ln m}{d \phi^2}\dot{\phi}\delta\phi,
\end{eqnarray}
\begin{equation}
\label{eq:theta}
    \dot{\theta}  + a H\theta =   \frac{d \ln m}{d \phi} k^{2} \delta \phi-\frac{d \ln m}{d \phi}\dot{\phi}\theta .
\end{equation}
where $\delta$ and $\theta \equiv \partial_i v^i$ are the density and velocity divergence perturbations of the dark matter, and $h$ is the metric trace perturbation in synchronous gauge.
The equations of motion for other components, including metric fluctuations, remain unchanged
(see, e.g.,~\cite{Ma:1995ey}).

These equations contain a number of notable features for CMB observables, which we return to in \S \ref{sec:pert}.  For large-scale structure, the sourcing of the scalar field fluctuations $\delta \phi$ from DM fluctuations $\delta$ is of particular interest. On small scales, where gradients dominate over both temporal derivatives and derivatives of the potential, the sourced scalar field fluctuation in this quasistatic approximation is given by
\begin{equation}
    \delta \phi ^{\rm (sourced)} \approx - \frac{a^2}{k^2}\frac{d\ln m}{d\phi}\rho_{\rm DM}\delta ,
\end{equation}
which acts as a slowly varying offset corresponding to the DM density-dependent minimum of the field oscillations.
This in turn sources a change in the DM momentum proportional to $\delta$, via Eq.~\ref{eq:theta}, in effect a dark ``fifth force'' \cite{McDonough:2021pdg} (see also related discussion in \cite{Bean:2007ny}).

In the case of an exponential coupling ${d\ln m}/{d\phi}=c/M_{\rm pl}$, the impact on DM can be described by an effective Newton's constant on small scales~\cite{McDonough:2021pdg},
\begin{equation}
\label{eq:GNeff}
    G_{\rm eff} = G_N(1+2c^2), \qquad ({\rm exp}) 
\end{equation}
leading to an enhanced growth of structure which significantly constrains the allowable range of $c$.  This may be contrasted with the case of the quadratic coupling of the tEDS model, which gives ${d\ln m}/{d\phi} = 2 g \phi /(M_{\rm pl}^2 + g \phi^2)\rightarrow 0$ at late times and
\begin{equation} 
    \lim_{\phi\rightarrow 0} G_{\rm eff} = G_N, \qquad ({\rm tEDS}).
\end{equation}
It follows that there is no enhanced growth of structure at late times from a dark fifth force for our tEDS model.

\section{$H_0$ Tension Data and Solutions}
\label{sec:data}

\begin{table*}
    \centering
    \begin{tabular}{c|c|c|c|c|c}
        \hline
        \hline
        Model & $\Lambda$CDM & EDE & tEDS($p$=4) & tEDS($p$=8) & tEDS($p$=16) \\
        \hline
        $100\theta_s$      & 1.04204 & 1.04129 & 1.04133 & 1.04112 & 1.04128  \\
        $\Omega_bh^2$      & 0.02254 & 0.02272 & 0.02285 & 0.02291 & 0.02293  \\
        $\Omega_ch^2$      & 0.1182  & 0.1319  & 0.1280  & 0.1280  & 0.1278   \\
        $\tau$             & 0.0595  & 0.0602  & 0.0559  & 0.0563  & 0.0549   \\
        $\ln (10^{10}A_s)$ & 3.052   & 3.075   & 3.056   & 3.055   & 3.055    \\
        $n_s$              & 0.9696  & 0.9887  & 0.9826  & 0.9825  & 0.9851   \\
        $\theta_i$         & -       & 2.768   & 1.208   & 1.103   & 1.056    \\
        $g$                & -       & -       & 0.076   & 0.041   & 0.017    \\
        $f/M_{\rm pl}$     & -       & 0.18    & 0.21    & 0.30    & 0.29     \\
        $V_0/{\rm eV}^4$   & -       & 0.050   & 2.15    & 1.96    & 1.75     \\
        \hline
        $f_{\rm EDE}$      & -       & 0.123   & 0.102   & 0.110   & 0.108    \\
        $\log_{10}z_c$     & -       & 3.57    & 3.87    & 3.85    & 3.84     \\
        $H_0$              & 68.24   & 71.90   & 70.46   & 70.72   & 70.77    \\
        $S_8$              & 0.8136  & 0.8437  & 0.8351  & 0.8305  & 0.8291   \\
        $f_{\rm EDE}/f_{\rm EDE}^{g=0}$ & -  & 1       & 0.31    & 0.32    & 0.38     \\
        \hline
        $\chi^2_{\rm TTTEEE}$ & 2346.1 & 2342.7 & 2347.4 & 2344.5  & 2343.8   \\
        $\Delta\chi^2_{\rm tot}$     & 0       & -17.4   & -10.1   & -14.6   & -14.5    \\
        \hline
        \hline
    \end{tabular}    
    \caption{
    Parameters of the best-fit models to the baseline datasets (CMB + BAO + SNe + $H_0$).  
    The first set of rows includes the fundamental model parameters, the second set the derived parameters, and the third set the goodness of fit, with $\Delta\chi^2_{\rm tot}$ relative to the $\Lambda$CDM model.
    }
    \label{tab:minimum}
\end{table*}

In order to assess the ability of our tEDS model to resolve the Hubble tension, we employ the following datasets: 

\begin{itemize}
    \item {\bf CMB}:  low-$\ell$ and high-$\ell$ {\it Planck} 2018 \cite{Planck2018likelihood,Aghanim:2018eyx,2018arXiv180706210P} [\texttt{Plik}] temperature and polarization power spectra (TT+TE+EE) and lensing potential power spectrum.
    
    \item {\bf BAO}: SDSS DR7 main galaxy sample~\cite{Ross:2014qpa}, 6dF galaxy survey~\cite{2011MNRAS.416.3017B}, and SDSS BOSS DR12 LOWZ+CMASS galaxy samples~\cite{Alam:2016hwk}.
    
    \item {\bf Supernovae}: Pantheon supernovae dataset of relative luminosity distances  \cite{Scolnic:2017caz}.
    
    \item ${\boldsymbol{H_0}}$: SH0ES 2019 Cepheid-Supernova  distance ladder measurement $H_0=74.03 \pm 1.42 \, {\rm km/s/Mpc}$ \cite{Riess:2019cxk}.\footnote{We use the SH0ES 2019 measurement to facilitate comparison with EDE results in the literature, whereas the most recent update gives $H_0=73.04 \pm 1.04 \, {\rm km/s/Mpc}$~\cite{Riess:2021jrx}.}
\end{itemize}

We perform the analyses of the tEDS model specified by Eqs.~\eqref{eq:mDM_quad} and \eqref{eq:V} using a modified version of CLASS\footnote{\url{http://class-code.net}} \cite{2011arXiv1104.2932L,2011JCAP...07..034B}.
In addition to the six standard $\Lambda$CDM parameters ($\theta_s$, $\Omega_bh^2$, $\Omega_ch^2$, $\tau$, $A_s$, $n_s$), the tEDS model includes three parameters that exist in the EDE model of Eq.~(\ref{eq:EDEpotential}) (potential scale $V_0$, field scale $f$, and initial field value $\phi_i= \theta_i f$)  and adds one parameter (the coupling $g$).  
Finally, the index $p$ determines the sharpness of the transition from the plateau in $V(\phi)$, which we take to be fixed to a sufficiently large value to provide a sharp transition.  

In the absence of well-motivated physical priors for these parameters and to establish a proof of principle parameter set that resolves the Hubble tension, we use  \texttt{Cobaya} \cite{torrado_lewis_2019} to find the best-fit model rather than sample the posterior distributions.  
Since \texttt{Cobaya minimize} can easily get stuck in a local minimum if the theory model is complicated (even with the covariance matrix from the MCMC chains), we develop an iterative minimizer routine that repeats \texttt{minimize} by setting the starting point as the last-round minimum until $\Delta\chi^2$ between the current and last-round minima is less than 0.1. Compared to the standard one-round minimizer, this iterative procedure can improve the result by $\Delta\chi^2\sim 3$.

These best-fit parameters and $\Delta \chi^2$ values for tEDS models with $p=4,8,16$ are given in Table~\ref{tab:minimum}, along with those of the best-fit $\Lambda$CDM and EDE models.  For example, for the $p=8$ tEDS model the fit  is better than $\Lambda$CDM by $\Delta\chi^2=-14.6$ with four additional free parameters, and gives $H_0=70.72\,\,{\rm km/s/Mpc}$. 
Notice that this model has a non-zero coupling $g=0.041$ and an initial field position on the plateau of the bare potential $\theta_i \approx 1.1$, but near its edge.  The fit is only marginally worse than EDE where $\Delta\chi^2 =-17.4$, with the additional parameter $g$ introduced to solve the coincidence problem rather than added to the EDE bare potential and optimized to improve its fit.  This should be borne in mind when assessing any model selection criteria.

Since the effective potential at finite $g$ evolves with redshift and so always dominates over the bare potential early on, to assess whether the coupling triggers the field to roll we compute the maximum EDE fraction for $g=0$, $f_{\rm EDE}^{g=0}$, with the same other parameters in the model.  In the limit that the coupling has no effect, the ratio $f_{\rm EDE}/f_{\rm EDE}^{g=0} \rightarrow 1$; whereas if the field fails to roll at all without the coupling such that EDE eventually dominates the expansion, $f_{\rm EDE}^{g=0}\rightarrow 1$ and so $f_{\rm EDE}/f_{\rm EDE}^{g=0} \rightarrow {f_{\rm EDE}}$.   
The best-fit $p=8$ model has  $f_{\rm EDE}/f_{\rm EDE}^{g=0}=0.31$, indicating that the roll is indeed triggered by the DM coupling. Thus, condition 2 in Sec.~\ref{sec:bg} is satisfied. Likewise, condition 1 is satisfied since $f_{\rm EDE}=0.11$ and $z_c=10^{3.85}$. 

\begin{figure}
    \centering
    \includegraphics[width=0.99\columnwidth]{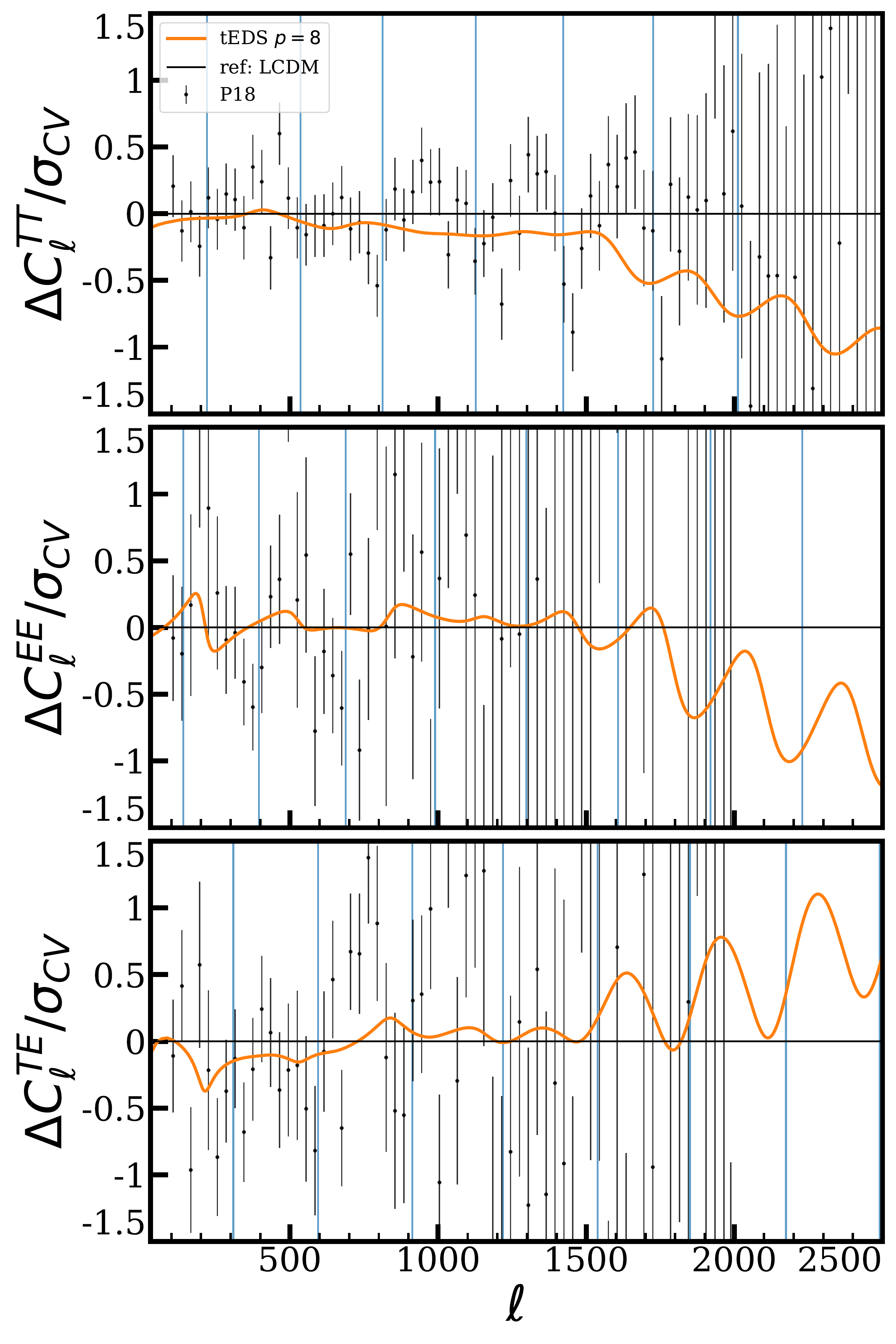}
    \caption{
    CMB TT, TE, and EE power spectra residuals (in units of the CV-limited error bar) for the $p=8$ best-fit model with respect to the best-fit $\Lambda$CDM model and the \emph{Planck} 2018 data.  Vertical lines represent the position of the acoustic peaks in this model.  
    Slowly varying residuals versus $\Lambda$CDM are absorbed by foreground parameters in the fit.
    }
    \label{fig:CMB-best}
\end{figure}

Finally, although the trigger mechanism for the background applies to a wide range of $\theta_i$, satisfying condition 3, the best-fit prefers a specific value $\theta_i\approx 1.1$.   We shall see in the next section that this value best flattens the CMB TT residuals relative to $\Lambda$CDM, especially for modes near the horizon at $z_c$, as shown in Fig.~\ref{fig:CMB-best}.   The residuals here are plotted relative to the best-fit $\Lambda$CDM model, expressed in units of the cosmic variance per multipole,
\begin{equation}
\sigma^{\rm CV}_\ell =
\begin{cases}
\sqrt{\frac{2}{2\ell+1}} C_\ell^{TT}, & {TT} \,;\\
\sqrt{\frac{1}{2\ell+1}}  \sqrt{ C_\ell^{TT} C_\ell^{EE} + (C_\ell^{TE})^2}, & {TE} \,;\\
\sqrt{\frac{2}{2\ell+1}}  C_\ell^{EE}, & {EE}. \, \\
\end{cases}
\end{equation}
Note that the smooth residuals, especially at high-$\ell$, are compensated by adjusting \emph{Planck} foreground parameters,\footnote{The main change is a small increase in $A_{\rm cib}^{217}$, the amplitude of the cosmic infrared background power spectrum at 217 GHz, which actually moves slightly closer to its reference value in the \emph{Planck} likelihood.}
whose effects are not shown in this figure since the data points are plotted assuming foreground parameters for $\Lambda$CDM.

In Table \ref{tab:minimum} we also show that the fits are only weakly sensitive to the sharpness of the transition. Compared to $p=8$, a sharper transition $p=16$ fits the data essentially equally as well, while a smoother transition $p=4$ is marginally worse.   Notice that the values of $f_{\rm EDE}$, $z_c$, and $f_{\rm EDE}/f_{\rm EDE}^{g=0}$ remain nearly the same.  Consequently we hereafter focus on the $p=8$ case. 

\section{Dynamical Balance in the CMB}
\label{sec:pert}

In the previous section we have shown that the tEDS model of Sec.~\ref{sec:bg} provides an excellent fit to the Hubble tension data, including the SH0ES $H_0$ measurement and {\it Planck} 2018 CMB power spectra. The fit is comparable to that in EDE whilst being free of the EDE coincidence problem in the background (see Fig.~\ref{fig:theta}).

On the other hand, although the trigger mechanism works to provide the desired energy density fraction $f_{\rm EDE}$ and the correct peak redshift $z_c$ to solve the Hubble tension for a wide range of initial field values $\theta_i>1$, the data prefer a specific value where the field gets a small but finite amount of its roll off the plateau from the bare potential, as monitored by $f_{\rm EDE}^{g=0}$, the EDE fraction with vanishing coupling $g$ (see Table~\ref{tab:minimum}).

The novel feature of the tEDS model is that although the field can successfully roll from a wide range of $\theta_i> 1$ to the edge of the plateau at $\theta=1$ around $z_c$, the kinetic energy of the field near $\theta=1$ then necessarily increases with $\theta_i$.   
This in turn generates field fluctuations which compete with those generated from falling off the plateau. The latter ones are largely independent of $\theta_i$.   
We shall see that it is a balance between these two effects that produces the $\theta_i$ preferred by the CMB data.

\begin{figure}[t]
    \centering
    \includegraphics[width=0.99\columnwidth]{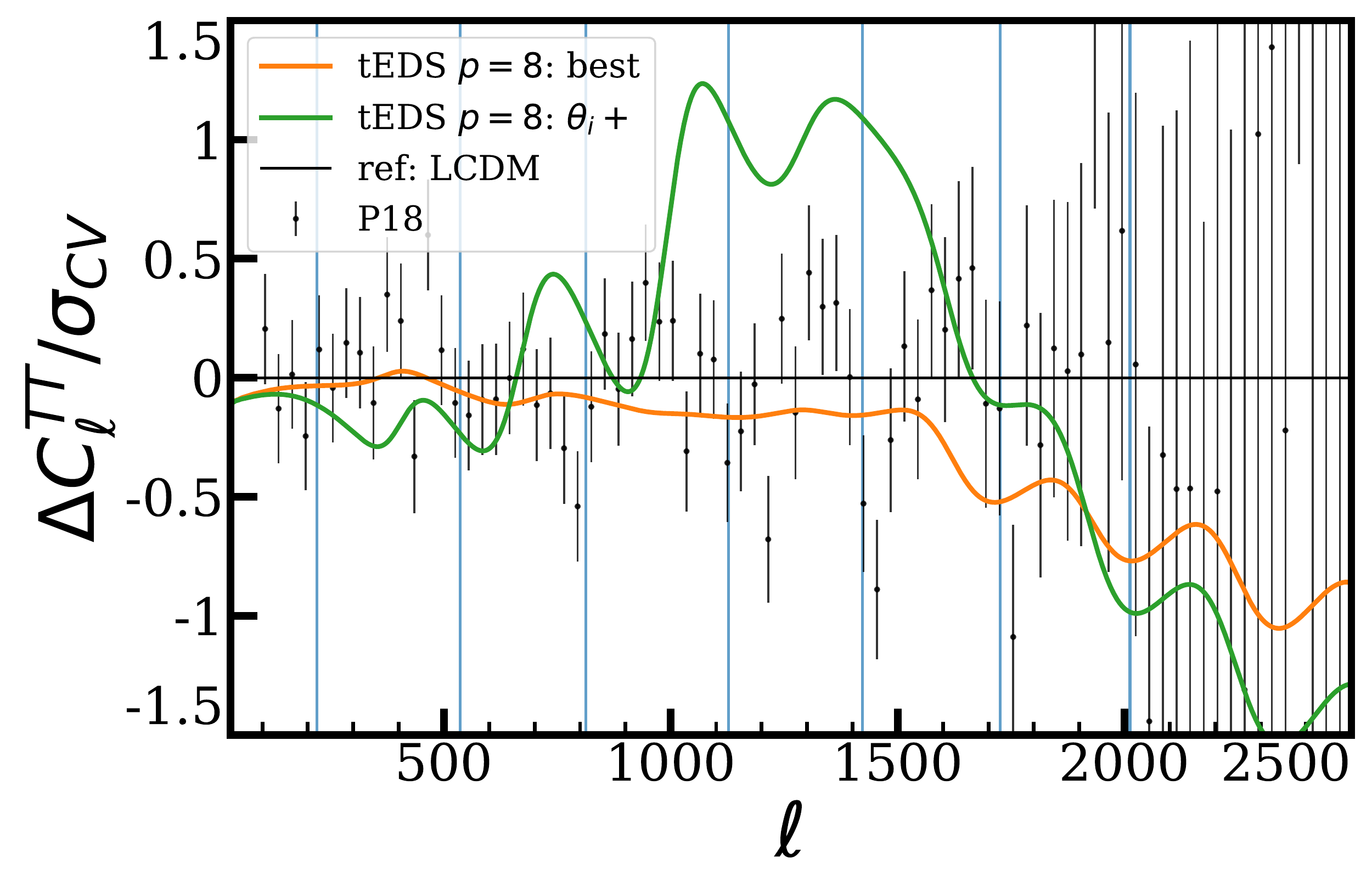}
    \caption{
    CMB TT power spectrum residuals (as in Fig.~\ref{fig:CMB-best}) for the $p=8$ best-fit model with initial field position $\theta_i=1.103$ and its variation with a higher $\theta_i=1.3$ and the same $z_c$ and $V(\phi)$.  Other aspects follow Fig.~\ref{fig:CMB-best}.
    }
    \label{fig:CMB-vary}
\end{figure}

\begin{figure*}
    \centering
    \includegraphics[width=1.99\columnwidth]{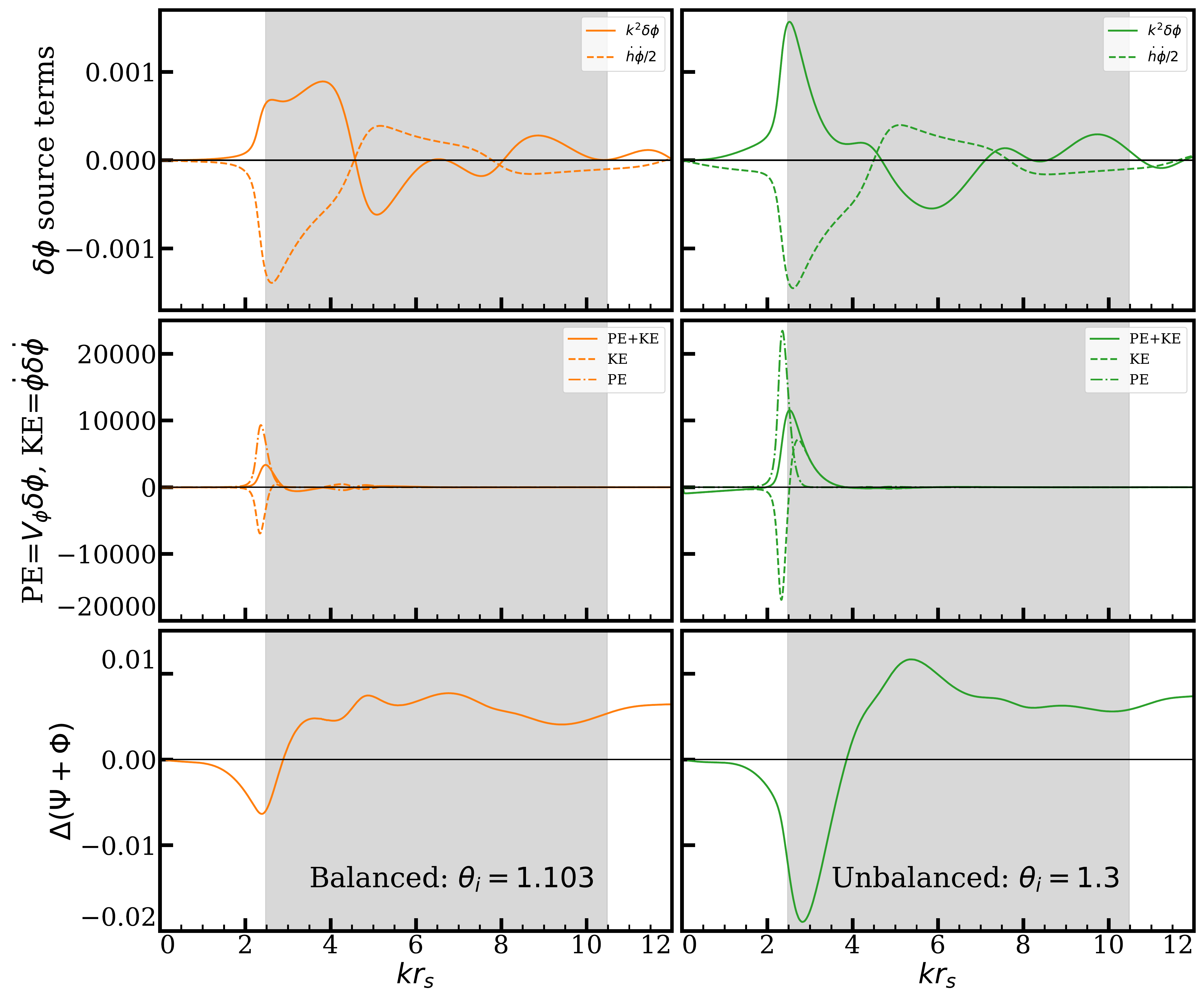}
    \caption{
    The evolution of the scalar field perturbation with its metric source (top panel), energy density fluctuations (middle panel), and the Weyl potential (bottom panel) of the horizon crossing mode at $z_c$ with $k=0.075$ Mpc$^{-1}$.
    The {\it left} panel shows the best-fit $p=8$ model with $\theta_i=1.103$, while the {\it right} shows an unbalanced case with a larger $\theta_i=1.3$, corresponding to the poorly fitting model in Fig.~\ref{fig:CMB-vary}.
    The Weyl potential is plotted relative to the best-fit $\Lambda$CDM model.  The region $z_{c}>z>z_*$ is shaded. 
    }
    \label{fig:deltaphi}
\end{figure*}

The consequence of this balance can be seen in the CMB residuals.
To illustrate this, in Fig.~\ref{fig:CMB-vary} we show the CMB TT spectra of the $p=8$ model for two examples: the best-fit model with $\theta_i=1.103$ and a variation to the best-fit with a larger value $\theta_i = 1.3$. The EDS coupling $g$ is adjusted to keep 
$z_c$ fixed with the same $V(\phi)$, leaving $f_{\rm EDE}=0.111$ nearly unchanged.

From Fig.~\ref{fig:CMB-vary} one may appreciate that the increase in $\theta_i$ induces a sizeable excess in the CMB TT spectrum in the multipole range $1000\lesssim\ell\lesssim 1500$, corresponding to modes that re-entered the horizon around $z_c$. This bump is not easily compensated by shifts in other cosmological or foreground parameters, and will remain even when the other parameters are re-optimized with $\theta_i$ held fixed to the larger value. 

We can trace the origin of this bump back to the behavior of the scalar field fluctuations $\delta\phi$ on the corresponding scales $k\approx 0.075$\,Mpc$^{-1}$.  In the absence of isocurvature initial conditions, the scalar field perturbations are predominantly sourced by the adiabatic metric perturbations, i.e., the term $\dot{h}\dot{\phi}$ appearing in Eq.~\eqref{eq:deltaphiddot}. The metric perturbation itself is relatively insensitive to the EDE, being predominantly sourced by the matter and radiation components of the universe. However, the larger $\theta_i$ model, while indistinguishable from any other $\theta_i$ at late times $(z\geq z_c)$, is distinguished by a larger velocity $\dot{\phi}$ at early times $(z<z_c)$. This is a consequence of satisfying the trigger condition in Eq.~\eqref{eq:release}, where for a fixed $V_{\rm eff}$, $\Delta \dot\phi \propto \Delta \theta_i$.  
In our example, we fix $z_c$ so that $g$ also varies somewhat with $\theta_i$.
Since $z_c$ corresponds roughly to the epoch at which $\theta=1$, the result is that $\dot\phi \propto (\theta_i-1)$ since by construction the field rolls to this position from its initial value in the time interval $t(z_c)$, as seen in the top panel of Fig.~\ref{fig:theta}. 
Approximating Eq.~\eqref{eq:deltaphiddot} on super-horizon scales and at early times $z<z_c$, we have
\begin{equation}
     \ddot{\delta\phi} \sim a H \dot{\delta \phi} \sim -\dot{h} \dot{\phi} .
\end{equation}
Integrating over time from the initial moment until $z_c$, and approximating $\dot{\phi}$ as a constant, we find $\delta \phi \propto (\theta_i-1)$, indicating an early-time growth of $\delta \phi$ that itself grows linearly with the initial condition $\theta_i$. 

In Fig.~\ref{fig:deltaphi} (top panels, solid lines) we compare the field fluctuations for the two different $\theta_i$ where the shaded region corresponds to $z_c > z> z_*$, where $z_*$ is the epoch of recombination when the acoustic CMB fluctuations are frozen in. 
For the time variable, we choose the comoving sound horizon of the photon-baryon fluid
\begin{equation}
    r_s(z) = \int_z^\infty \frac{c_s(\Tilde{z}) d\Tilde{z}}{H(\Tilde{z})},
\end{equation}
where $c_s$ is the photon-baryon sound speed to highlight the relevant epoch for the acoustic oscillations. Here we have weighted $\delta\phi$ by $k^2$ for ease of comparison to the metric sources (dashed lines). Notice the strong dependence on $\theta_i$ for $z_c< z$.    

At $z_c$, when the EDE scalar rolls down its potential and the EDE begins to decay, there is a dramatic increase in the field velocity $\dot{\phi}$ due to the bare potential. The sharp growth in $\dot{\phi}$ imparts a `kick' on the perturbations $\delta \phi$, again coming from the $\dot{\phi} \dot{h}$ source term in Eq.~\eqref{eq:deltaphiddot}, but now largely independent of $\theta_i$ and at a time where the initial kick has already evolved to a different phase of its now subhorizon evolution.
In Fig.~\ref{fig:theta} (top panels, dashed lines), we can see that the metric sources at $z<z_c$ are nearly indistinguishable between the two $\theta_i$ values.

The combination of these two kicks to the field fluctuations, one dependent on $\theta_i$ and the other independent, imply that there is a special value where the kicks are balanced so that $\delta\phi$ is nearly constant in time at the crucial epoch for driving CMB acoustic oscillations around $k r_s(z) \approx \pi$.

The impact on the CMB from the EDS is through its gravitational effect via the Weyl potential $\Psi+\Phi$ (see, e.g.,~\cite{Lin:2018nxe}), which in turn reflects the energy density fluctuations carried by the field.  In Fig.~\ref{fig:deltaphi} (middle panel), we separate these fluctuations into the potential (PE) and kinetic (KE) energy contributions.  Because of the rapid conversion of potential to kinetic energy, the two begin at $z>z_c$ nearly equal and opposite and allow for a much faster evolution of the total than usually allowed by an equation of state $|w_\phi|\le 1$ \cite{Mortonson:2009qq} as the kinetic energy redshifts away.  Correspondingly the Weyl potential evolves rapidly and drives the CMB acoustic oscillations (bottom panel). 
This feature is not present in the EDE model and depends strongly on $\theta_i$.   

After $z_c$ the potential energy corresponding to a given field fluctuation sharply declines as the field falls off the plateau and leaves the dominant energy density fluctuation as contributed by kinetic energy.  In this regime, the counterbalancing metric kicks that leave the field fluctuation nearly constant for $\theta_i=1.103$ reduce the energy density fluctuation and hence the impact on the Weyl potential, especially for the variations on the $k \Delta r_s\sim \pi$ time scale that drive CMB acoustic oscillations.   It is this combination that is responsible for the minimization of CMB residuals in Fig.~\ref{fig:CMB-best}.   
For $1\lesssim \theta_i \lesssim 1.1$ this imbalance shifts in the other direction, though not as dramatically as for a large increase in $\theta_i$.  

In the case of $\theta_i\lesssim 1$, the slope of the bare potential dominates the initial roll and the model effectively reduces to the $g=0$ case with a power-law potential, which has limited ability to solve the Hubble tension (see, e.g.,~\cite{Agrawal:2019lmo}).

In summary, the best-fit tEDS model reflects a balance between the dynamics of the trigger in the background, which is largely insensitive to the initial field value $\theta_i$ by design, and the dynamics of the perturbations where there is a competition between metric kicks whose balance depends on $\theta_i$.   In this sense, the resolution of the Hubble tension requires more than just the resolution of the coincidence problem of the background, but is nonetheless successfully achieved within the tEDS model.

\section{$S_8$ Tension: tEDS vs.~EDE} 
\label{sec:LSS}

\begin{figure}[t]
    \centering
    \includegraphics[width=0.99\columnwidth]{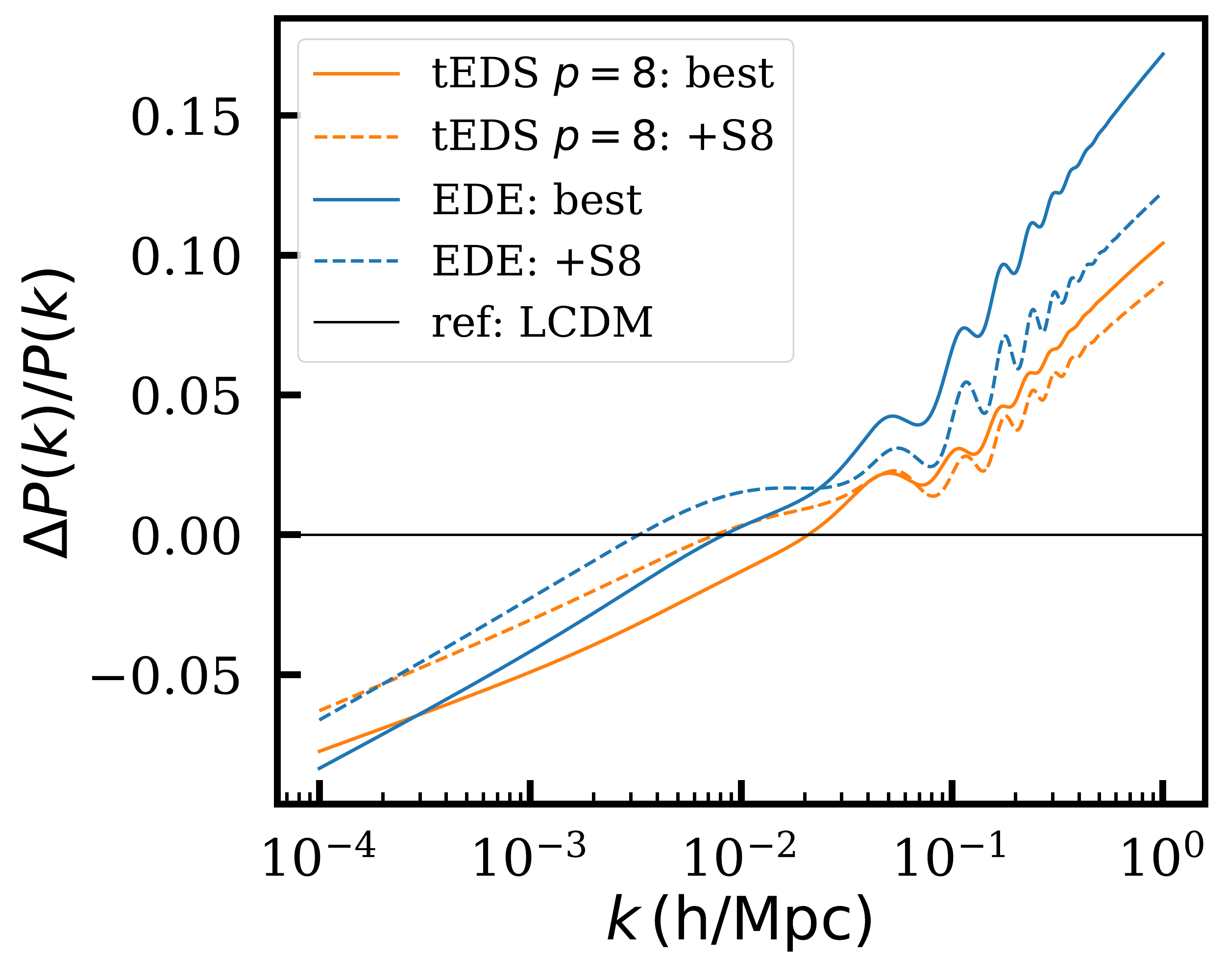}
    \caption{
    Matter power spectra for the tEDS $p=8$ and EDE models with respect to the best-fit $\Lambda$CDM model. Both best-fit models for the baseline (``best") and $+S_8$ datasets are shown.
    }
    \label{fig:Pk}
\end{figure}

The tEDS model performs comparably to EDE in resolving the Hubble tension, namely, in the fit to the baseline data set comprised of \emph{Planck} 2018 CMB primary anisotropies and CMB lensing, BOSS BAO, Pantheon SNIa, and SH0ES. The $\chi^2$ difference between the maximum-likelihood tEDS ($p=8$) and EDE models is $\Delta \chi^2 _{\rm tot,tEDS - EDE} = + 2.8$ (see Table~\ref{tab:minimum}), indicating that the coincidence-resolving tEDS model is a slightly worse fit to cosmological data than EDE. However, the two best-fit models are also distinguished by their $S_8$ values: $S_8 = 0.8291$ and $S_8 = 0.8436$ in tEDS and EDE, respectively. This difference reduces the relative tension between the tEDS model and large-scale structure data: the EDE model is in $4.0\sigma$ tension with the Dark Energy Survey Year-3 (DES-Y3) measurement $S_8 =0.776 \pm 0.017 $ \cite{DES:2021wwk}, compared to only $3.1\sigma$ in tEDS.  The matter power spectra for the best-fit tEDS and EDE models to the baseline data sets are shown with respect to the best-fit $\Lambda$CDM model in Fig.~\ref{fig:Pk} (solid lines). 
Notice that the $n_s$ increase compared to $\Lambda$CDM is somewhat reduced in tEDS compared to EDE, which partially helps lower $S_8$ in tEDS as does the lower $\Omega_c h^2$.

Of course, these best-fit models for the Hubble tension would have their parameters readjusted to lower $S_8$ if the DES-Y3 data were included, but the trade-offs may differ between EDE and tEDS. 

Motivated by this potential difference, we supplement the baseline datasets with additional large-scale structure data from the DES-Y3 analysis \cite{DES:2021wwk}:
\begin{itemize}
    \item {\bf DES-Y3}: Dark Energy Survey Year-3 \cite{DES:2021wwk} weak lensing and galaxy clustering data, namely, galaxy-galaxy, shear-shear, and galaxy-shear two-point correlation functions (``3$\times$2-point''), implemented as a Gaussian constraint on $S_8\equiv \sigma_8 (\Omega_{\rm m} / 0.3)^{0.5}$ corresponding to the DES-Y3 measurement $S_8 = 0.776 \pm 0.017$.
\end{itemize}
We repeat our analysis and search for maximum likelihood parameters again using \texttt{Cobaya} \cite{torrado_lewis_2019}.  Note that the approximation of treating the full DES 3$\times$2-point likelihood as an effective Gaussian prior on $S_8$ was validated for $\Lambda$CDM and EDE in Ref.~\cite{Hill:2020osr} (in that case, for DES-Y1), and thus we adopt this approach here as well.

\begin{table}
    \centering
    \begin{tabular}{c|c|c}
        \hline
        \hline
        Model & EDE & tEDS($p$=8) \\
        \hline
        $100\theta_s$      & 1.04139 & 1.04112   \\
        $\Omega_bh^2$      & 0.02281 & 0.02296   \\
        $\Omega_ch^2$      & 0.1287  & 0.1273     \\
        $\tau$             & 0.0581  & 0.0565     \\
        $\ln(10^{10}A_s)$ & 3.065   & 3.054      \\
        $n_s$              & 0.9894  & 0.9843     \\
        $\theta_i$         & 2.763   & 1.103      \\
        $g$                & -       & 0.039      \\
        $f/M_{\rm pl}$     & 0.17    & 0.30       \\
        $V_0/{\rm eV}^4$   & 0.040   & 1.74      \\
        \hline
        $f_{\rm EDE}$      & 0.108   & 0.112       \\
        $\log_{10}z_c$     & 3.56    & 3.83        \\
        $H_0$              & 71.96   & 71.21      \\
        $S_8$              & 0.8236  & 0.8200     \\
        \hline
        $\chi^2_{\rm TTTEEE}$ & 2345.2 & 2346.9     \\
        $\chi^2_{\rm tot}-\chi^2_{\rm tot,EDE}$     & 0       & +1.2      \\
        \hline
        \hline
    \end{tabular}    
    \caption{
    Parameters of the best-fit models to the baseline + $S_8$ datasets. 
    }
    \label{tab:minimumS8}
\end{table}

The maximum-likelihood model parameters when DES-Y3 data is included are given in Table~\ref{tab:minimumS8}. The $\chi^2$ difference between the models is now $\Delta \chi^2 _{\rm tot,tEDS - EDE}=+1.2$, indicating that the tEDS model becomes even closer to EDE in goodness of fit. Interestingly, the values for both $H_0$ and $S_8$ of the best fits become closer as well, with $H_0$ remaining nearly unchanged in EDE and actually increasing in tEDS.  
In Fig.~\ref{fig:Pk} (dashed lines), we also show the matter power spectra for these $+S_8$ data best-fit models.

While these new best-fit cases would apparently reduce the $S_8$ and $H_0$ tensions simultaneously, the overall fit to the baseline data set without DES is, of course, slightly worse.  This comes primarily from the CMB, where in both cases $\Delta\chi^2_{\rm TTTEEE}\approx +2.5$.   The poorer fit and parameter trends can be explained from the residual freedom to adjust the $\Lambda$CDM parameters themselves. 
At the expense of the \emph{Planck} 2018 fit for $\Omega_c h^2$, $\Lambda$CDM models follow the CMB angular diameter distance degeneracy leading to  $S_8 \propto h^{-2.3}$ (e.g.,~\cite{Secco:2022kqg}).   The best-fit EDE and tEDS models can then exploit this scaling to lower $S_8$ without lowering $H_0$ and compromise between the competing demands of the various datasets.

\section{Discussion}
\label{sec:discussion}

In this work we have formalized and studied in detail the coincidence problem of EDE models, working primarily in the context an {\it early dark sector}, in which the mass of the dark matter particle is dependent on the EDE scalar field. The coincidence problem is naturally a statement pertaining to the background evolution: why is the decay of the background energy density in the EDE field anchored to the epoch when the background energy densities of matter and radiation are equal?  From this, one may enumerate the requirements for a coincidence-free EDS model: 1.\ Realize the desired EDE-like dynamics ($f_{\rm EDE}\sim 10\%$ and $z_c \sim 10^{3.5}$), with 2.\  dark matter triggered decay, and 3.\ no fine-tuning of initial conditions. We have demonstrated that a {\it trigger EDS (tEDS)} model with a plateau-like potential $V(\phi)$ and a quadratic coupling to dark matter, $\Delta m_{\rm DM} \propto (\phi/M_{\rm pl})^2$, can naturally satisfy these conditions on the background evolution.

We find that the tEDS model performs comparably to EDE when fit to a combined data set comprised of \emph{Planck} 2018 CMB temperature and polarization anisotropies and CMB lensing, BOSS BAO, Pantheon SNIa, and the SH0ES cosmic distance ladder measurement of $H_0$. This fit becomes even closer and essentially indistinguishably as good if $S_8$ constraints are included from DES-Y3, achieving $H_0 =71.2$ km/s/Mpc.

However, amongst the trigger solutions that provide the same background evolution, fitting the CMB anisotropies associated with perturbations that cross the horizon around $z_c$ selects a specific range of initial conditions for the background value of the EDE scalar. 
We find that the optimal tEDS model is one wherein there is a dynamical balance of effects in the evolution of field fluctuations induced by the trigger and by the bare potential.  In our choice of coupling and bare potential, this balance is enforced by the CMB data rather than built into the model itself, which suggests that future refinements of the basic trigger mechanism may be possible.

Another novel feature of this optimal parameter choice is that $z_c$ is slightly higher than $z_{\rm eq}$, which leads to observable differences from the best-fit $\Lambda$CDM and EDE models that are distinguishable at higher multipole moments than are well-constrained by the \emph{Planck} 2018 data.  These predictions can be tested in the near future.  Of course as with the $S_8$ tension and data, the best-fit parameters of tEDS may also change with new data sets, requiring a full analysis of parameter posteriors to assess model performance rather than the best-fit approach here.

More generally, the assessment of EDE and EDE-like models is sensitive to the data sets that are included or excluded in the analysis as well as priors on their parameters. This is consistent with past work, such as Ref.~\cite{Hill:2020osr} where EDE was analyzed using different data set combinations and parameter priors (e.g., priors on model parameters versus on derived parameters $f_{\rm EDE}$, $z_c$) with dramatically different outcomes. See also Ref.~\cite{Herold:2021ksg} for related discussion.
With this in mind, an interesting next step for EDS will be to investigate constraints on the parameters of this scenario with the ACT DR4 \cite{ACT:2020gnv,ACT:2020frw} and
SPT-3G 2018 \cite{SPT-3G:2021eoc,Balkenhol:2022rvc} data, particularly in light of the mild ACT preference for a non-zero EDE component \cite{Lin:2020jcb,Hill:2021yec,Smith:2022hwi}, analogous to the EDE analysis in \cite{LaPosta:2021pgm}, as well as the BOSS full-shape data, e.g., utilizing an effective field theory-based large-scale structure likelihood, as done for EDE in \cite{Ivanov:2020ril,DAmico:2020ods}. We leave these and other interesting directions to future work.


\acknowledgments
We thank Leah Jenks, Austin Joyce, Tanvi Karwal, Hayden Lee, Vivian Poulin, Marco Raveri, and Tristan Smith for useful discussions and comments.
M.X.L. and W.H.\ 
were supported by U.S.\ Dept.\ of Energy contract DE-FG02-13ER41958 and the Simons Foundation. 
M.X.L. was also supported in part by funds provided by the Center for Particle Cosmology.
E.M.\ was supported in part by a Discovery Grant from the National Sciences and Engineering Research Council of Canada.
J.C.H.\ acknowledges support from NSF grant AST-2108536, NASA grant 21-ATP21-0129, DOE grant DE-SC00233966, the Sloan Foundation, and the Simons Foundation.
Computing resources were provided by the University of Chicago Research Computing Center through the Kavli Institute for Cosmological Physics at the University of Chicago.
\bibliography{EDS-refs}

\begin{thebibliography}{62}%
\makeatletter
\providecommand \@ifxundefined [1]{%
 \@ifx{#1\undefined}
}%
\providecommand \@ifnum [1]{%
 \ifnum #1\expandafter \@firstoftwo
 \else \expandafter \@secondoftwo
 \fi
}%
\providecommand \@ifx [1]{%
 \ifx #1\expandafter \@firstoftwo
 \else \expandafter \@secondoftwo
 \fi
}%
\providecommand \natexlab [1]{#1}%
\providecommand \enquote  [1]{``#1''}%
\providecommand \bibnamefont  [1]{#1}%
\providecommand \bibfnamefont [1]{#1}%
\providecommand \citenamefont [1]{#1}%
\providecommand \href@noop [0]{\@secondoftwo}%
\providecommand \href [0]{\begingroup \@sanitize@url \@href}%
\providecommand \@href[1]{\@@startlink{#1}\@@href}%
\providecommand \@@href[1]{\endgroup#1\@@endlink}%
\providecommand \@sanitize@url [0]{\catcode `\\12\catcode `\$12\catcode
  `\&12\catcode `\#12\catcode `\^12\catcode `\_12\catcode `\%12\relax}%
\providecommand \@@startlink[1]{}%
\providecommand \@@endlink[0]{}%
\providecommand \url  [0]{\begingroup\@sanitize@url \@url }%
\providecommand \@url [1]{\endgroup\@href {#1}{\urlprefix }}%
\providecommand \urlprefix  [0]{URL }%
\providecommand \Eprint [0]{\href }%
\providecommand \doibase [0]{http://dx.doi.org/}%
\providecommand \selectlanguage [0]{\@gobble}%
\providecommand \bibinfo  [0]{\@secondoftwo}%
\providecommand \bibfield  [0]{\@secondoftwo}%
\providecommand \translation [1]{[#1]}%
\providecommand \BibitemOpen [0]{}%
\providecommand \bibitemStop [0]{}%
\providecommand \bibitemNoStop [0]{.\EOS\space}%
\providecommand \EOS [0]{\spacefactor3000\relax}%
\providecommand \BibitemShut  [1]{\csname bibitem#1\endcsname}%
\let\auto@bib@innerbib\@empty
\bibitem [{\citenamefont {Aghanim}\ \emph {et~al.}(2018)\citenamefont {Aghanim}
  \emph {et~al.}}]{Aghanim:2018eyx}%
  \BibitemOpen
  \bibfield  {author} {\bibinfo {author} {\bibfnamefont {N.}~\bibnamefont
  {Aghanim}} \emph {et~al.} (\bibinfo {collaboration} {Planck}),\ }\href@noop
  {} {\  (\bibinfo {year} {2018})},\ \Eprint {http://arxiv.org/abs/1807.06209}
  {arXiv:1807.06209 [astro-ph.CO]} \BibitemShut {NoStop}%
\bibitem [{\citenamefont {Riess}\ \emph {et~al.}(2021)\citenamefont {Riess}
  \emph {et~al.}}]{Riess:2021jrx}%
  \BibitemOpen
  \bibfield  {author} {\bibinfo {author} {\bibfnamefont {A.~G.}\ \bibnamefont
  {Riess}} \emph {et~al.},\ }\href@noop {} {\  (\bibinfo {year} {2021})},\
  \Eprint {http://arxiv.org/abs/2112.04510} {arXiv:2112.04510 [astro-ph.CO]}
  \BibitemShut {NoStop}%
\bibitem [{\citenamefont {Verde}\ \emph {et~al.}(2019)\citenamefont {Verde},
  \citenamefont {Treu},\ and\ \citenamefont {Riess}}]{Verde:2019ivm}%
  \BibitemOpen
  \bibfield  {author} {\bibinfo {author} {\bibfnamefont {L.}~\bibnamefont
  {Verde}}, \bibinfo {author} {\bibfnamefont {T.}~\bibnamefont {Treu}}, \ and\
  \bibinfo {author} {\bibfnamefont {A.~G.}\ \bibnamefont {Riess}},\ }in\ \href
  {\doibase 10.1038/s41550-019-0902-0} {\emph {\bibinfo {booktitle} {{Nature
  Astronomy 2019}}}}\ (\bibinfo {year} {2019})\ \Eprint
  {http://arxiv.org/abs/1907.10625} {arXiv:1907.10625 [astro-ph.CO]}
  \BibitemShut {NoStop}%
\bibitem [{\citenamefont {Kamionkowski}\ and\ \citenamefont
  {Riess}(2022)}]{Kamionkowski:2022pkx}%
  \BibitemOpen
  \bibfield  {author} {\bibinfo {author} {\bibfnamefont {M.}~\bibnamefont
  {Kamionkowski}}\ and\ \bibinfo {author} {\bibfnamefont {A.~G.}\ \bibnamefont
  {Riess}},\ }\href@noop {} {\  (\bibinfo {year} {2022})},\ \Eprint
  {http://arxiv.org/abs/2211.04492} {arXiv:2211.04492 [astro-ph.CO]}
  \BibitemShut {NoStop}%
\bibitem [{\citenamefont {Freedman}\ \emph {et~al.}(2019)\citenamefont
  {Freedman} \emph {et~al.}}]{Freedman:2019jwv}%
  \BibitemOpen
  \bibfield  {author} {\bibinfo {author} {\bibfnamefont {W.~L.}\ \bibnamefont
  {Freedman}} \emph {et~al.},\ }\href {\doibase 10.3847/1538-4357/ab2f73} {\
  (\bibinfo {year} {2019}),\ 10.3847/1538-4357/ab2f73},\ \Eprint
  {http://arxiv.org/abs/1907.05922} {arXiv:1907.05922 [astro-ph.CO]}
  \BibitemShut {NoStop}%
\bibitem [{\citenamefont {Freedman}(2021)}]{Freedman:2021ahq}%
  \BibitemOpen
  \bibfield  {author} {\bibinfo {author} {\bibfnamefont {W.~L.}\ \bibnamefont
  {Freedman}},\ }\href {\doibase 10.3847/1538-4357/ac0e95} {\bibfield
  {journal} {\bibinfo  {journal} {Astrophys. J.}\ }\textbf {\bibinfo {volume}
  {919}},\ \bibinfo {pages} {16} (\bibinfo {year} {2021})},\ \Eprint
  {http://arxiv.org/abs/2106.15656} {arXiv:2106.15656 [astro-ph.CO]}
  \BibitemShut {NoStop}%
\bibitem [{\citenamefont {Karwal}\ and\ \citenamefont
  {Kamionkowski}(2016)}]{Karwal:2016vyq}%
  \BibitemOpen
  \bibfield  {author} {\bibinfo {author} {\bibfnamefont {T.}~\bibnamefont
  {Karwal}}\ and\ \bibinfo {author} {\bibfnamefont {M.}~\bibnamefont
  {Kamionkowski}},\ }\href {\doibase 10.1103/PhysRevD.94.103523} {\bibfield
  {journal} {\bibinfo  {journal} {Phys. Rev. D}\ }\textbf {\bibinfo {volume}
  {94}},\ \bibinfo {pages} {103523} (\bibinfo {year} {2016})},\ \Eprint
  {http://arxiv.org/abs/1608.01309} {arXiv:1608.01309 [astro-ph.CO]}
  \BibitemShut {NoStop}%
\bibitem [{\citenamefont {Poulin}\ \emph {et~al.}(2019)\citenamefont {Poulin},
  \citenamefont {Smith}, \citenamefont {Karwal},\ and\ \citenamefont
  {Kamionkowski}}]{Poulin:2018cxd}%
  \BibitemOpen
  \bibfield  {author} {\bibinfo {author} {\bibfnamefont {V.}~\bibnamefont
  {Poulin}}, \bibinfo {author} {\bibfnamefont {T.~L.}\ \bibnamefont {Smith}},
  \bibinfo {author} {\bibfnamefont {T.}~\bibnamefont {Karwal}}, \ and\ \bibinfo
  {author} {\bibfnamefont {M.}~\bibnamefont {Kamionkowski}},\ }\href {\doibase
  10.1103/PhysRevLett.122.221301} {\bibfield  {journal} {\bibinfo  {journal}
  {Phys. Rev. Lett.}\ }\textbf {\bibinfo {volume} {122}},\ \bibinfo {pages}
  {221301} (\bibinfo {year} {2019})},\ \Eprint
  {http://arxiv.org/abs/1811.04083} {arXiv:1811.04083 [astro-ph.CO]}
  \BibitemShut {NoStop}%
\bibitem [{\citenamefont {Agrawal}\ \emph {et~al.}(2019)\citenamefont
  {Agrawal}, \citenamefont {Cyr-Racine}, \citenamefont {Pinner},\ and\
  \citenamefont {Randall}}]{Agrawal:2019lmo}%
  \BibitemOpen
  \bibfield  {author} {\bibinfo {author} {\bibfnamefont {P.}~\bibnamefont
  {Agrawal}}, \bibinfo {author} {\bibfnamefont {F.-Y.}\ \bibnamefont
  {Cyr-Racine}}, \bibinfo {author} {\bibfnamefont {D.}~\bibnamefont {Pinner}},
  \ and\ \bibinfo {author} {\bibfnamefont {L.}~\bibnamefont {Randall}},\
  }\href@noop {} {\  (\bibinfo {year} {2019})},\ \Eprint
  {http://arxiv.org/abs/1904.01016} {arXiv:1904.01016 [astro-ph.CO]}
  \BibitemShut {NoStop}%
\bibitem [{\citenamefont {Lin}\ \emph {et~al.}(2019{\natexlab{a}})\citenamefont
  {Lin}, \citenamefont {Benevento}, \citenamefont {Hu},\ and\ \citenamefont
  {Raveri}}]{Lin:2019qug}%
  \BibitemOpen
  \bibfield  {author} {\bibinfo {author} {\bibfnamefont {M.-X.}\ \bibnamefont
  {Lin}}, \bibinfo {author} {\bibfnamefont {G.}~\bibnamefont {Benevento}},
  \bibinfo {author} {\bibfnamefont {W.}~\bibnamefont {Hu}}, \ and\ \bibinfo
  {author} {\bibfnamefont {M.}~\bibnamefont {Raveri}},\ }\href {\doibase
  10.1103/PhysRevD.100.063542} {\bibfield  {journal} {\bibinfo  {journal}
  {Phys. Rev. D}\ }\textbf {\bibinfo {volume} {100}},\ \bibinfo {pages}
  {063542} (\bibinfo {year} {2019}{\natexlab{a}})},\ \Eprint
  {http://arxiv.org/abs/1905.12618} {arXiv:1905.12618 [astro-ph.CO]}
  \BibitemShut {NoStop}%
\bibitem [{\citenamefont {Smith}\ \emph {et~al.}(2020)\citenamefont {Smith},
  \citenamefont {Poulin},\ and\ \citenamefont {Amin}}]{Smith:2019ihp}%
  \BibitemOpen
  \bibfield  {author} {\bibinfo {author} {\bibfnamefont {T.~L.}\ \bibnamefont
  {Smith}}, \bibinfo {author} {\bibfnamefont {V.}~\bibnamefont {Poulin}}, \
  and\ \bibinfo {author} {\bibfnamefont {M.~A.}\ \bibnamefont {Amin}},\ }\href
  {\doibase 10.1103/PhysRevD.101.063523} {\bibfield  {journal} {\bibinfo
  {journal} {Phys. Rev. D}\ }\textbf {\bibinfo {volume} {101}},\ \bibinfo
  {pages} {063523} (\bibinfo {year} {2020})},\ \Eprint
  {http://arxiv.org/abs/1908.06995} {arXiv:1908.06995 [astro-ph.CO]}
  \BibitemShut {NoStop}%
\bibitem [{\citenamefont {Alexander}\ and\ \citenamefont
  {McDonough}(2019)}]{Alexander:2019rsc}%
  \BibitemOpen
  \bibfield  {author} {\bibinfo {author} {\bibfnamefont {S.}~\bibnamefont
  {Alexander}}\ and\ \bibinfo {author} {\bibfnamefont {E.}~\bibnamefont
  {McDonough}},\ }\href {\doibase 10.1016/j.physletb.2019.134830} {\bibfield
  {journal} {\bibinfo  {journal} {Phys. Lett.}\ }\textbf {\bibinfo {volume}
  {B797}},\ \bibinfo {pages} {134830} (\bibinfo {year} {2019})},\ \Eprint
  {http://arxiv.org/abs/1904.08912} {arXiv:1904.08912 [astro-ph.CO]}
  \BibitemShut {NoStop}%
\bibitem [{\citenamefont {Alexander}\ \emph {et~al.}(2022)\citenamefont
  {Alexander}, \citenamefont {Bernardo},\ and\ \citenamefont
  {Toomey}}]{Alexander:2022own}%
  \BibitemOpen
  \bibfield  {author} {\bibinfo {author} {\bibfnamefont {S.}~\bibnamefont
  {Alexander}}, \bibinfo {author} {\bibfnamefont {H.}~\bibnamefont {Bernardo}},
  \ and\ \bibinfo {author} {\bibfnamefont {M.~W.}\ \bibnamefont {Toomey}},\
  }\href@noop {} {\  (\bibinfo {year} {2022})},\ \Eprint
  {http://arxiv.org/abs/2207.13086} {arXiv:2207.13086 [astro-ph.CO]}
  \BibitemShut {NoStop}%
\bibitem [{\citenamefont {Knox}\ and\ \citenamefont
  {Millea}(2020)}]{Knox:2019rjx}%
  \BibitemOpen
  \bibfield  {author} {\bibinfo {author} {\bibfnamefont {L.}~\bibnamefont
  {Knox}}\ and\ \bibinfo {author} {\bibfnamefont {M.}~\bibnamefont {Millea}},\
  }\href {\doibase 10.1103/PhysRevD.101.043533} {\bibfield  {journal} {\bibinfo
   {journal} {Phys. Rev. D}\ }\textbf {\bibinfo {volume} {101}},\ \bibinfo
  {pages} {043533} (\bibinfo {year} {2020})},\ \Eprint
  {http://arxiv.org/abs/1908.03663} {arXiv:1908.03663 [astro-ph.CO]}
  \BibitemShut {NoStop}%
\bibitem [{\citenamefont {Sabla}\ and\ \citenamefont
  {Caldwell}(2021)}]{Sabla:2021nfy}%
  \BibitemOpen
  \bibfield  {author} {\bibinfo {author} {\bibfnamefont {V.~I.}\ \bibnamefont
  {Sabla}}\ and\ \bibinfo {author} {\bibfnamefont {R.~R.}\ \bibnamefont
  {Caldwell}},\ }\href {\doibase 10.1103/PhysRevD.103.103506} {\bibfield
  {journal} {\bibinfo  {journal} {Phys. Rev. D}\ }\textbf {\bibinfo {volume}
  {103}},\ \bibinfo {pages} {103506} (\bibinfo {year} {2021})},\ \Eprint
  {http://arxiv.org/abs/2103.04999} {arXiv:2103.04999 [astro-ph.CO]}
  \BibitemShut {NoStop}%
\bibitem [{\citenamefont {Sakstein}\ and\ \citenamefont
  {Trodden}(2019)}]{Sakstein:2019fmf}%
  \BibitemOpen
  \bibfield  {author} {\bibinfo {author} {\bibfnamefont {J.}~\bibnamefont
  {Sakstein}}\ and\ \bibinfo {author} {\bibfnamefont {M.}~\bibnamefont
  {Trodden}},\ }\href@noop {} {\  (\bibinfo {year} {2019})},\ \Eprint
  {http://arxiv.org/abs/1911.11760} {arXiv:1911.11760 [astro-ph.CO]}
  \BibitemShut {NoStop}%
\bibitem [{\citenamefont {Carrillo~Gonz\'alez}\ \emph
  {et~al.}(2021)\citenamefont {Carrillo~Gonz\'alez}, \citenamefont {Liang},
  \citenamefont {Sakstein},\ and\ \citenamefont
  {Trodden}}]{CarrilloGonzalez:2020oac}%
  \BibitemOpen
  \bibfield  {author} {\bibinfo {author} {\bibfnamefont {M.}~\bibnamefont
  {Carrillo~Gonz\'alez}}, \bibinfo {author} {\bibfnamefont {Q.}~\bibnamefont
  {Liang}}, \bibinfo {author} {\bibfnamefont {J.}~\bibnamefont {Sakstein}}, \
  and\ \bibinfo {author} {\bibfnamefont {M.}~\bibnamefont {Trodden}},\ }\href
  {\doibase 10.1088/1475-7516/2021/04/063} {\bibfield  {journal} {\bibinfo
  {journal} {JCAP}\ }\textbf {\bibinfo {volume} {04}},\ \bibinfo {pages} {063}
  (\bibinfo {year} {2021})},\ \Eprint {http://arxiv.org/abs/2011.09895}
  {arXiv:2011.09895 [astro-ph.CO]} \BibitemShut {NoStop}%
\bibitem [{\citenamefont {Karwal}\ \emph {et~al.}(2022)\citenamefont {Karwal},
  \citenamefont {Raveri}, \citenamefont {Jain}, \citenamefont {Khoury},\ and\
  \citenamefont {Trodden}}]{Karwal:2021vpk}%
  \BibitemOpen
  \bibfield  {author} {\bibinfo {author} {\bibfnamefont {T.}~\bibnamefont
  {Karwal}}, \bibinfo {author} {\bibfnamefont {M.}~\bibnamefont {Raveri}},
  \bibinfo {author} {\bibfnamefont {B.}~\bibnamefont {Jain}}, \bibinfo {author}
  {\bibfnamefont {J.}~\bibnamefont {Khoury}}, \ and\ \bibinfo {author}
  {\bibfnamefont {M.}~\bibnamefont {Trodden}},\ }\href {\doibase
  10.1103/PhysRevD.105.063535} {\bibfield  {journal} {\bibinfo  {journal}
  {Phys. Rev. D}\ }\textbf {\bibinfo {volume} {105}},\ \bibinfo {pages}
  {063535} (\bibinfo {year} {2022})},\ \Eprint
  {http://arxiv.org/abs/2106.13290} {arXiv:2106.13290 [astro-ph.CO]}
  \BibitemShut {NoStop}%
\bibitem [{\citenamefont {McDonough}\ \emph {et~al.}(2022)\citenamefont
  {McDonough}, \citenamefont {Lin}, \citenamefont {Hill}, \citenamefont {Hu},\
  and\ \citenamefont {Zhou}}]{McDonough:2021pdg}%
  \BibitemOpen
  \bibfield  {author} {\bibinfo {author} {\bibfnamefont {E.}~\bibnamefont
  {McDonough}}, \bibinfo {author} {\bibfnamefont {M.-X.}\ \bibnamefont {Lin}},
  \bibinfo {author} {\bibfnamefont {J.~C.}\ \bibnamefont {Hill}}, \bibinfo
  {author} {\bibfnamefont {W.}~\bibnamefont {Hu}}, \ and\ \bibinfo {author}
  {\bibfnamefont {S.}~\bibnamefont {Zhou}},\ }\href {\doibase
  10.1103/PhysRevD.106.043525} {\bibfield  {journal} {\bibinfo  {journal}
  {Phys. Rev. D}\ }\textbf {\bibinfo {volume} {106}},\ \bibinfo {pages}
  {043525} (\bibinfo {year} {2022})},\ \Eprint
  {http://arxiv.org/abs/2112.09128} {arXiv:2112.09128 [astro-ph.CO]}
  \BibitemShut {NoStop}%
\bibitem [{\citenamefont {Hill}\ \emph {et~al.}(2020)\citenamefont {Hill},
  \citenamefont {McDonough}, \citenamefont {Toomey},\ and\ \citenamefont
  {Alexander}}]{Hill:2020osr}%
  \BibitemOpen
  \bibfield  {author} {\bibinfo {author} {\bibfnamefont {J.~C.}\ \bibnamefont
  {Hill}}, \bibinfo {author} {\bibfnamefont {E.}~\bibnamefont {McDonough}},
  \bibinfo {author} {\bibfnamefont {M.~W.}\ \bibnamefont {Toomey}}, \ and\
  \bibinfo {author} {\bibfnamefont {S.}~\bibnamefont {Alexander}},\ }\href
  {\doibase 10.1103/PhysRevD.102.043507} {\bibfield  {journal} {\bibinfo
  {journal} {Phys. Rev. D}\ }\textbf {\bibinfo {volume} {102}},\ \bibinfo
  {pages} {043507} (\bibinfo {year} {2020})},\ \Eprint
  {http://arxiv.org/abs/2003.07355} {arXiv:2003.07355 [astro-ph.CO]}
  \BibitemShut {NoStop}%
\bibitem [{\citenamefont {Ivanov}\ \emph {et~al.}(2020)\citenamefont {Ivanov},
  \citenamefont {McDonough}, \citenamefont {Hill}, \citenamefont {Simonovi\'c},
  \citenamefont {Toomey}, \citenamefont {Alexander},\ and\ \citenamefont
  {Zaldarriaga}}]{Ivanov:2020ril}%
  \BibitemOpen
  \bibfield  {author} {\bibinfo {author} {\bibfnamefont {M.~M.}\ \bibnamefont
  {Ivanov}}, \bibinfo {author} {\bibfnamefont {E.}~\bibnamefont {McDonough}},
  \bibinfo {author} {\bibfnamefont {J.~C.}\ \bibnamefont {Hill}}, \bibinfo
  {author} {\bibfnamefont {M.}~\bibnamefont {Simonovi\'c}}, \bibinfo {author}
  {\bibfnamefont {M.~W.}\ \bibnamefont {Toomey}}, \bibinfo {author}
  {\bibfnamefont {S.}~\bibnamefont {Alexander}}, \ and\ \bibinfo {author}
  {\bibfnamefont {M.}~\bibnamefont {Zaldarriaga}},\ }\href {\doibase
  10.1103/PhysRevD.102.103502} {\bibfield  {journal} {\bibinfo  {journal}
  {Phys. Rev. D}\ }\textbf {\bibinfo {volume} {102}},\ \bibinfo {pages}
  {103502} (\bibinfo {year} {2020})},\ \Eprint
  {http://arxiv.org/abs/2006.11235} {arXiv:2006.11235 [astro-ph.CO]}
  \BibitemShut {NoStop}%
\bibitem [{\citenamefont {D'Amico}\ \emph {et~al.}(2021)\citenamefont
  {D'Amico}, \citenamefont {Senatore}, \citenamefont {Zhang},\ and\
  \citenamefont {Zheng}}]{DAmico:2020ods}%
  \BibitemOpen
  \bibfield  {author} {\bibinfo {author} {\bibfnamefont {G.}~\bibnamefont
  {D'Amico}}, \bibinfo {author} {\bibfnamefont {L.}~\bibnamefont {Senatore}},
  \bibinfo {author} {\bibfnamefont {P.}~\bibnamefont {Zhang}}, \ and\ \bibinfo
  {author} {\bibfnamefont {H.}~\bibnamefont {Zheng}},\ }\href {\doibase
  10.1088/1475-7516/2021/05/072} {\bibfield  {journal} {\bibinfo  {journal}
  {JCAP}\ }\textbf {\bibinfo {volume} {05}},\ \bibinfo {pages} {072} (\bibinfo
  {year} {2021})},\ \Eprint {http://arxiv.org/abs/2006.12420} {arXiv:2006.12420
  [astro-ph.CO]} \BibitemShut {NoStop}%
\bibitem [{\citenamefont {Jedamzik}\ \emph {et~al.}(2020)\citenamefont
  {Jedamzik}, \citenamefont {Pogosian},\ and\ \citenamefont
  {Zhao}}]{Jedamzik:2020zmd}%
  \BibitemOpen
  \bibfield  {author} {\bibinfo {author} {\bibfnamefont {K.}~\bibnamefont
  {Jedamzik}}, \bibinfo {author} {\bibfnamefont {L.}~\bibnamefont {Pogosian}},
  \ and\ \bibinfo {author} {\bibfnamefont {G.-B.}\ \bibnamefont {Zhao}},\
  }\href@noop {} {\  (\bibinfo {year} {2020})},\ \Eprint
  {http://arxiv.org/abs/2010.04158} {arXiv:2010.04158 [astro-ph.CO]}
  \BibitemShut {NoStop}%
\bibitem [{\citenamefont {Lin}\ \emph {et~al.}(2021)\citenamefont {Lin},
  \citenamefont {Chen},\ and\ \citenamefont {Mack}}]{Lin:2021sfs}%
  \BibitemOpen
  \bibfield  {author} {\bibinfo {author} {\bibfnamefont {W.}~\bibnamefont
  {Lin}}, \bibinfo {author} {\bibfnamefont {X.}~\bibnamefont {Chen}}, \ and\
  \bibinfo {author} {\bibfnamefont {K.~J.}\ \bibnamefont {Mack}},\ }\href
  {\doibase 10.3847/1538-4357/ac12cf} {\bibfield  {journal} {\bibinfo
  {journal} {Astrophys. J.}\ }\textbf {\bibinfo {volume} {920}},\ \bibinfo
  {pages} {159} (\bibinfo {year} {2021})},\ \Eprint
  {http://arxiv.org/abs/2102.05701} {arXiv:2102.05701 [astro-ph.CO]}
  \BibitemShut {NoStop}%
\bibitem [{\citenamefont {Smith}\ \emph {et~al.}(2021)\citenamefont {Smith},
  \citenamefont {Poulin}, \citenamefont {Bernal}, \citenamefont {Boddy},
  \citenamefont {Kamionkowski},\ and\ \citenamefont {Murgia}}]{Smith:2020rxx}%
  \BibitemOpen
  \bibfield  {author} {\bibinfo {author} {\bibfnamefont {T.~L.}\ \bibnamefont
  {Smith}}, \bibinfo {author} {\bibfnamefont {V.}~\bibnamefont {Poulin}},
  \bibinfo {author} {\bibfnamefont {J.~L.}\ \bibnamefont {Bernal}}, \bibinfo
  {author} {\bibfnamefont {K.~K.}\ \bibnamefont {Boddy}}, \bibinfo {author}
  {\bibfnamefont {M.}~\bibnamefont {Kamionkowski}}, \ and\ \bibinfo {author}
  {\bibfnamefont {R.}~\bibnamefont {Murgia}},\ }\href {\doibase
  10.1103/PhysRevD.103.123542} {\bibfield  {journal} {\bibinfo  {journal}
  {Phys. Rev. D}\ }\textbf {\bibinfo {volume} {103}},\ \bibinfo {pages}
  {123542} (\bibinfo {year} {2021})},\ \Eprint
  {http://arxiv.org/abs/2009.10740} {arXiv:2009.10740 [astro-ph.CO]}
  \BibitemShut {NoStop}%
\bibitem [{\citenamefont {Murgia}\ \emph {et~al.}(2021)\citenamefont {Murgia},
  \citenamefont {Abell\'an},\ and\ \citenamefont {Poulin}}]{Murgia:2020ryi}%
  \BibitemOpen
  \bibfield  {author} {\bibinfo {author} {\bibfnamefont {R.}~\bibnamefont
  {Murgia}}, \bibinfo {author} {\bibfnamefont {G.~F.}\ \bibnamefont
  {Abell\'an}}, \ and\ \bibinfo {author} {\bibfnamefont {V.}~\bibnamefont
  {Poulin}},\ }\href {\doibase 10.1103/PhysRevD.103.063502} {\bibfield
  {journal} {\bibinfo  {journal} {Phys. Rev. D}\ }\textbf {\bibinfo {volume}
  {103}},\ \bibinfo {pages} {063502} (\bibinfo {year} {2021})},\ \Eprint
  {http://arxiv.org/abs/2009.10733} {arXiv:2009.10733 [astro-ph.CO]}
  \BibitemShut {NoStop}%
\bibitem [{\citenamefont {Simon}\ \emph {et~al.}(2022)\citenamefont {Simon},
  \citenamefont {Zhang}, \citenamefont {Poulin},\ and\ \citenamefont
  {Smith}}]{Simon:2022adh}%
  \BibitemOpen
  \bibfield  {author} {\bibinfo {author} {\bibfnamefont {T.}~\bibnamefont
  {Simon}}, \bibinfo {author} {\bibfnamefont {P.}~\bibnamefont {Zhang}},
  \bibinfo {author} {\bibfnamefont {V.}~\bibnamefont {Poulin}}, \ and\ \bibinfo
  {author} {\bibfnamefont {T.~L.}\ \bibnamefont {Smith}},\ }\href@noop {} {\
  (\bibinfo {year} {2022})},\ \Eprint {http://arxiv.org/abs/2208.05930}
  {arXiv:2208.05930 [astro-ph.CO]} \BibitemShut {NoStop}%
\bibitem [{\citenamefont {Herold}\ \emph {et~al.}(2022)\citenamefont {Herold},
  \citenamefont {Ferreira},\ and\ \citenamefont {Komatsu}}]{Herold:2021ksg}%
  \BibitemOpen
  \bibfield  {author} {\bibinfo {author} {\bibfnamefont {L.}~\bibnamefont
  {Herold}}, \bibinfo {author} {\bibfnamefont {E.~G.~M.}\ \bibnamefont
  {Ferreira}}, \ and\ \bibinfo {author} {\bibfnamefont {E.}~\bibnamefont
  {Komatsu}},\ }\href {\doibase 10.3847/2041-8213/ac63a3} {\bibfield  {journal}
  {\bibinfo  {journal} {Astrophys. J. Lett.}\ }\textbf {\bibinfo {volume}
  {929}},\ \bibinfo {pages} {L16} (\bibinfo {year} {2022})},\ \Eprint
  {http://arxiv.org/abs/2112.12140} {arXiv:2112.12140 [astro-ph.CO]}
  \BibitemShut {NoStop}%
\bibitem [{\citenamefont {Baumann}\ \emph {et~al.}(2006)\citenamefont
  {Baumann}, \citenamefont {Dymarsky}, \citenamefont {Klebanov}, \citenamefont
  {Maldacena}, \citenamefont {McAllister},\ and\ \citenamefont
  {Murugan}}]{Baumann:2006th}%
  \BibitemOpen
  \bibfield  {author} {\bibinfo {author} {\bibfnamefont {D.}~\bibnamefont
  {Baumann}}, \bibinfo {author} {\bibfnamefont {A.}~\bibnamefont {Dymarsky}},
  \bibinfo {author} {\bibfnamefont {I.~R.}\ \bibnamefont {Klebanov}}, \bibinfo
  {author} {\bibfnamefont {J.~M.}\ \bibnamefont {Maldacena}}, \bibinfo {author}
  {\bibfnamefont {L.~P.}\ \bibnamefont {McAllister}}, \ and\ \bibinfo {author}
  {\bibfnamefont {A.}~\bibnamefont {Murugan}},\ }\href {\doibase
  10.1088/1126-6708/2006/11/031} {\bibfield  {journal} {\bibinfo  {journal}
  {JHEP}\ }\textbf {\bibinfo {volume} {11}},\ \bibinfo {pages} {031} (\bibinfo
  {year} {2006})},\ \Eprint {http://arxiv.org/abs/hep-th/0607050}
  {arXiv:hep-th/0607050} \BibitemShut {NoStop}%
\bibitem [{\citenamefont {Ruehle}\ and\ \citenamefont
  {Wieck}(2017)}]{Ruehle:2017one}%
  \BibitemOpen
  \bibfield  {author} {\bibinfo {author} {\bibfnamefont {F.}~\bibnamefont
  {Ruehle}}\ and\ \bibinfo {author} {\bibfnamefont {C.}~\bibnamefont {Wieck}},\
  }\href {\doibase 10.1016/j.physletb.2017.03.072} {\bibfield  {journal}
  {\bibinfo  {journal} {Phys. Lett. B}\ }\textbf {\bibinfo {volume} {769}},\
  \bibinfo {pages} {289} (\bibinfo {year} {2017})},\ \Eprint
  {http://arxiv.org/abs/1702.00420} {arXiv:1702.00420 [hep-th]} \BibitemShut
  {NoStop}%
\bibitem [{\citenamefont {Dong}\ \emph {et~al.}(2011)\citenamefont {Dong},
  \citenamefont {Horn}, \citenamefont {Silverstein},\ and\ \citenamefont
  {Westphal}}]{Dong:2010in}%
  \BibitemOpen
  \bibfield  {author} {\bibinfo {author} {\bibfnamefont {X.}~\bibnamefont
  {Dong}}, \bibinfo {author} {\bibfnamefont {B.}~\bibnamefont {Horn}}, \bibinfo
  {author} {\bibfnamefont {E.}~\bibnamefont {Silverstein}}, \ and\ \bibinfo
  {author} {\bibfnamefont {A.}~\bibnamefont {Westphal}},\ }\href {\doibase
  10.1103/PhysRevD.84.026011} {\bibfield  {journal} {\bibinfo  {journal} {Phys.
  Rev. D}\ }\textbf {\bibinfo {volume} {84}},\ \bibinfo {pages} {026011}
  (\bibinfo {year} {2011})},\ \Eprint {http://arxiv.org/abs/1011.4521}
  {arXiv:1011.4521 [hep-th]} \BibitemShut {NoStop}%
\bibitem [{\citenamefont {McAllister}\ \emph {et~al.}(2010)\citenamefont
  {McAllister}, \citenamefont {Silverstein},\ and\ \citenamefont
  {Westphal}}]{McAllister:2008hb}%
  \BibitemOpen
  \bibfield  {author} {\bibinfo {author} {\bibfnamefont {L.}~\bibnamefont
  {McAllister}}, \bibinfo {author} {\bibfnamefont {E.}~\bibnamefont
  {Silverstein}}, \ and\ \bibinfo {author} {\bibfnamefont {A.}~\bibnamefont
  {Westphal}},\ }\href {\doibase 10.1103/PhysRevD.82.046003} {\bibfield
  {journal} {\bibinfo  {journal} {Phys. Rev. D}\ }\textbf {\bibinfo {volume}
  {82}},\ \bibinfo {pages} {046003} (\bibinfo {year} {2010})},\ \Eprint
  {http://arxiv.org/abs/0808.0706} {arXiv:0808.0706 [hep-th]} \BibitemShut
  {NoStop}%
\bibitem [{\citenamefont {Silverstein}\ and\ \citenamefont
  {Westphal}(2008)}]{Silverstein:2008sg}%
  \BibitemOpen
  \bibfield  {author} {\bibinfo {author} {\bibfnamefont {E.}~\bibnamefont
  {Silverstein}}\ and\ \bibinfo {author} {\bibfnamefont {A.}~\bibnamefont
  {Westphal}},\ }\href {\doibase 10.1103/PhysRevD.78.106003} {\bibfield
  {journal} {\bibinfo  {journal} {Phys. Rev. D}\ }\textbf {\bibinfo {volume}
  {78}},\ \bibinfo {pages} {106003} (\bibinfo {year} {2008})},\ \Eprint
  {http://arxiv.org/abs/0803.3085} {arXiv:0803.3085 [hep-th]} \BibitemShut
  {NoStop}%
\bibitem [{\citenamefont {McAllister}\ \emph {et~al.}(2014)\citenamefont
  {McAllister}, \citenamefont {Silverstein}, \citenamefont {Westphal},\ and\
  \citenamefont {Wrase}}]{McAllister:2014mpa}%
  \BibitemOpen
  \bibfield  {author} {\bibinfo {author} {\bibfnamefont {L.}~\bibnamefont
  {McAllister}}, \bibinfo {author} {\bibfnamefont {E.}~\bibnamefont
  {Silverstein}}, \bibinfo {author} {\bibfnamefont {A.}~\bibnamefont
  {Westphal}}, \ and\ \bibinfo {author} {\bibfnamefont {T.}~\bibnamefont
  {Wrase}},\ }\href {\doibase 10.1007/JHEP09(2014)123} {\bibfield  {journal}
  {\bibinfo  {journal} {JHEP}\ }\textbf {\bibinfo {volume} {09}},\ \bibinfo
  {pages} {123} (\bibinfo {year} {2014})},\ \Eprint
  {http://arxiv.org/abs/1405.3652} {arXiv:1405.3652 [hep-th]} \BibitemShut
  {NoStop}%
\bibitem [{\citenamefont {Amin}\ \emph {et~al.}(2012)\citenamefont {Amin},
  \citenamefont {Easther}, \citenamefont {Finkel}, \citenamefont {Flauger},\
  and\ \citenamefont {Hertzberg}}]{Amin:2011hj}%
  \BibitemOpen
  \bibfield  {author} {\bibinfo {author} {\bibfnamefont {M.~A.}\ \bibnamefont
  {Amin}}, \bibinfo {author} {\bibfnamefont {R.}~\bibnamefont {Easther}},
  \bibinfo {author} {\bibfnamefont {H.}~\bibnamefont {Finkel}}, \bibinfo
  {author} {\bibfnamefont {R.}~\bibnamefont {Flauger}}, \ and\ \bibinfo
  {author} {\bibfnamefont {M.~P.}\ \bibnamefont {Hertzberg}},\ }\href {\doibase
  10.1103/PhysRevLett.108.241302} {\bibfield  {journal} {\bibinfo  {journal}
  {Phys. Rev. Lett.}\ }\textbf {\bibinfo {volume} {108}},\ \bibinfo {pages}
  {241302} (\bibinfo {year} {2012})},\ \Eprint {http://arxiv.org/abs/1106.3335}
  {arXiv:1106.3335 [astro-ph.CO]} \BibitemShut {NoStop}%
\bibitem [{\citenamefont {Lozanov}\ and\ \citenamefont
  {Amin}(2017)}]{Lozanov:2016hid}%
  \BibitemOpen
  \bibfield  {author} {\bibinfo {author} {\bibfnamefont {K.~D.}\ \bibnamefont
  {Lozanov}}\ and\ \bibinfo {author} {\bibfnamefont {M.~A.}\ \bibnamefont
  {Amin}},\ }\href {\doibase 10.1103/PhysRevLett.119.061301} {\bibfield
  {journal} {\bibinfo  {journal} {Phys. Rev. Lett.}\ }\textbf {\bibinfo
  {volume} {119}},\ \bibinfo {pages} {061301} (\bibinfo {year} {2017})},\
  \Eprint {http://arxiv.org/abs/1608.01213} {arXiv:1608.01213 [astro-ph.CO]}
  \BibitemShut {NoStop}%
\bibitem [{\citenamefont {Lozanov}\ and\ \citenamefont
  {Amin}(2018)}]{Lozanov:2017hjm}%
  \BibitemOpen
  \bibfield  {author} {\bibinfo {author} {\bibfnamefont {K.~D.}\ \bibnamefont
  {Lozanov}}\ and\ \bibinfo {author} {\bibfnamefont {M.~A.}\ \bibnamefont
  {Amin}},\ }\href {\doibase 10.1103/PhysRevD.97.023533} {\bibfield  {journal}
  {\bibinfo  {journal} {Phys. Rev. D}\ }\textbf {\bibinfo {volume} {97}},\
  \bibinfo {pages} {023533} (\bibinfo {year} {2018})},\ \Eprint
  {http://arxiv.org/abs/1710.06851} {arXiv:1710.06851 [astro-ph.CO]}
  \BibitemShut {NoStop}%
\bibitem [{\citenamefont {Blumenhagen}\ \emph {et~al.}(2015)\citenamefont
  {Blumenhagen}, \citenamefont {Font}, \citenamefont {Fuchs}, \citenamefont
  {Herschmann},\ and\ \citenamefont {Plauschinn}}]{Blumenhagen:2015qda}%
  \BibitemOpen
  \bibfield  {author} {\bibinfo {author} {\bibfnamefont {R.}~\bibnamefont
  {Blumenhagen}}, \bibinfo {author} {\bibfnamefont {A.}~\bibnamefont {Font}},
  \bibinfo {author} {\bibfnamefont {M.}~\bibnamefont {Fuchs}}, \bibinfo
  {author} {\bibfnamefont {D.}~\bibnamefont {Herschmann}}, \ and\ \bibinfo
  {author} {\bibfnamefont {E.}~\bibnamefont {Plauschinn}},\ }\href {\doibase
  10.1016/j.physletb.2015.05.001} {\bibfield  {journal} {\bibinfo  {journal}
  {Phys. Lett. B}\ }\textbf {\bibinfo {volume} {746}},\ \bibinfo {pages} {217}
  (\bibinfo {year} {2015})},\ \Eprint {http://arxiv.org/abs/1503.01607}
  {arXiv:1503.01607 [hep-th]} \BibitemShut {NoStop}%
\bibitem [{\citenamefont {Ma}\ and\ \citenamefont
  {Bertschinger}(1995)}]{Ma:1995ey}%
  \BibitemOpen
  \bibfield  {author} {\bibinfo {author} {\bibfnamefont {C.-P.}\ \bibnamefont
  {Ma}}\ and\ \bibinfo {author} {\bibfnamefont {E.}~\bibnamefont
  {Bertschinger}},\ }\href {\doibase 10.1086/176550} {\bibfield  {journal}
  {\bibinfo  {journal} {Astrophys. J.}\ }\textbf {\bibinfo {volume} {455}},\
  \bibinfo {pages} {7} (\bibinfo {year} {1995})},\ \Eprint
  {http://arxiv.org/abs/astro-ph/9506072} {arXiv:astro-ph/9506072} \BibitemShut
  {NoStop}%
\bibitem [{\citenamefont {Bean}\ \emph {et~al.}(2008)\citenamefont {Bean},
  \citenamefont {Flanagan},\ and\ \citenamefont {Trodden}}]{Bean:2007ny}%
  \BibitemOpen
  \bibfield  {author} {\bibinfo {author} {\bibfnamefont {R.}~\bibnamefont
  {Bean}}, \bibinfo {author} {\bibfnamefont {E.~E.}\ \bibnamefont {Flanagan}},
  \ and\ \bibinfo {author} {\bibfnamefont {M.}~\bibnamefont {Trodden}},\ }\href
  {\doibase 10.1103/PhysRevD.78.023009} {\bibfield  {journal} {\bibinfo
  {journal} {Phys. Rev. D}\ }\textbf {\bibinfo {volume} {78}},\ \bibinfo
  {pages} {023009} (\bibinfo {year} {2008})},\ \Eprint
  {http://arxiv.org/abs/0709.1128} {arXiv:0709.1128 [astro-ph]} \BibitemShut
  {NoStop}%
\bibitem [{\citenamefont {{Planck
  Collaboration}}(2019)}]{Planck2018likelihood}%
  \BibitemOpen
  \bibfield  {author} {\bibinfo {author} {\bibnamefont {{Planck
  Collaboration}}},\ }\href@noop {} {\bibfield  {journal} {\bibinfo  {journal}
  {arXiv e-prints}\ ,\ \bibinfo {eid} {arXiv:1907.12875}} (\bibinfo {year}
  {2019})},\ \Eprint {http://arxiv.org/abs/1907.12875} {arXiv:1907.12875
  [astro-ph.CO]} \BibitemShut {NoStop}%
\bibitem [{\citenamefont {{Planck Collaboration}}(2018)}]{2018arXiv180706210P}%
  \BibitemOpen
  \bibfield  {author} {\bibinfo {author} {\bibnamefont {{Planck
  Collaboration}}},\ }\href@noop {} {\bibfield  {journal} {\bibinfo  {journal}
  {arXiv e-prints}\ ,\ \bibinfo {eid} {arXiv:1807.06210}} (\bibinfo {year}
  {2018})},\ \Eprint {http://arxiv.org/abs/1807.06210} {arXiv:1807.06210
  [astro-ph.CO]} \BibitemShut {NoStop}%
\bibitem [{\citenamefont {Ross}\ \emph {et~al.}(2015)\citenamefont {Ross},
  \citenamefont {Samushia}, \citenamefont {Howlett}, \citenamefont {Percival},
  \citenamefont {Burden},\ and\ \citenamefont {Manera}}]{Ross:2014qpa}%
  \BibitemOpen
  \bibfield  {author} {\bibinfo {author} {\bibfnamefont {A.~J.}\ \bibnamefont
  {Ross}}, \bibinfo {author} {\bibfnamefont {L.}~\bibnamefont {Samushia}},
  \bibinfo {author} {\bibfnamefont {C.}~\bibnamefont {Howlett}}, \bibinfo
  {author} {\bibfnamefont {W.~J.}\ \bibnamefont {Percival}}, \bibinfo {author}
  {\bibfnamefont {A.}~\bibnamefont {Burden}}, \ and\ \bibinfo {author}
  {\bibfnamefont {M.}~\bibnamefont {Manera}},\ }\href {\doibase
  10.1093/mnras/stv154} {\bibfield  {journal} {\bibinfo  {journal} {Mon. Not.
  Roy. Astron. Soc.}\ }\textbf {\bibinfo {volume} {449}},\ \bibinfo {pages}
  {835} (\bibinfo {year} {2015})},\ \Eprint {http://arxiv.org/abs/1409.3242}
  {arXiv:1409.3242 [astro-ph.CO]} \BibitemShut {NoStop}%
\bibitem [{\citenamefont {{Beutler}}\ \emph {et~al.}(2011)\citenamefont
  {{Beutler}}, \citenamefont {{Blake}}, \citenamefont {{Colless}},
  \citenamefont {{Jones}}, \citenamefont {{Staveley-Smith}}, \citenamefont
  {{Campbell}}, \citenamefont {{Parker}}, \citenamefont {{Saunders}},\ and\
  \citenamefont {{Watson}}}]{2011MNRAS.416.3017B}%
  \BibitemOpen
  \bibfield  {author} {\bibinfo {author} {\bibfnamefont {F.}~\bibnamefont
  {{Beutler}}}, \bibinfo {author} {\bibfnamefont {C.}~\bibnamefont {{Blake}}},
  \bibinfo {author} {\bibfnamefont {M.}~\bibnamefont {{Colless}}}, \bibinfo
  {author} {\bibfnamefont {D.~H.}\ \bibnamefont {{Jones}}}, \bibinfo {author}
  {\bibfnamefont {L.}~\bibnamefont {{Staveley-Smith}}}, \bibinfo {author}
  {\bibfnamefont {L.}~\bibnamefont {{Campbell}}}, \bibinfo {author}
  {\bibfnamefont {Q.}~\bibnamefont {{Parker}}}, \bibinfo {author}
  {\bibfnamefont {W.}~\bibnamefont {{Saunders}}}, \ and\ \bibinfo {author}
  {\bibfnamefont {F.}~\bibnamefont {{Watson}}},\ }\href {\doibase
  10.1111/j.1365-2966.2011.19250.x} {\bibfield  {journal} {\bibinfo  {journal}
  {\mnras}\ }\textbf {\bibinfo {volume} {416}},\ \bibinfo {pages} {3017}
  (\bibinfo {year} {2011})},\ \Eprint {http://arxiv.org/abs/1106.3366}
  {arXiv:1106.3366 [astro-ph.CO]} \BibitemShut {NoStop}%
\bibitem [{\citenamefont {Alam}\ \emph {et~al.}(2017)\citenamefont {Alam} \emph
  {et~al.}}]{Alam:2016hwk}%
  \BibitemOpen
  \bibfield  {author} {\bibinfo {author} {\bibfnamefont {S.}~\bibnamefont
  {Alam}} \emph {et~al.} (\bibinfo {collaboration} {BOSS}),\ }\href {\doibase
  10.1093/mnras/stx721} {\bibfield  {journal} {\bibinfo  {journal} {Mon. Not.
  Roy. Astron. Soc.}\ }\textbf {\bibinfo {volume} {470}},\ \bibinfo {pages}
  {2617} (\bibinfo {year} {2017})},\ \Eprint {http://arxiv.org/abs/1607.03155}
  {arXiv:1607.03155 [astro-ph.CO]} \BibitemShut {NoStop}%
\bibitem [{\citenamefont {Scolnic}\ \emph {et~al.}(2018)\citenamefont {Scolnic}
  \emph {et~al.}}]{Scolnic:2017caz}%
  \BibitemOpen
  \bibfield  {author} {\bibinfo {author} {\bibfnamefont {D.~M.}\ \bibnamefont
  {Scolnic}} \emph {et~al.},\ }\href {\doibase 10.3847/1538-4357/aab9bb}
  {\bibfield  {journal} {\bibinfo  {journal} {Astrophys. J.}\ }\textbf
  {\bibinfo {volume} {859}},\ \bibinfo {pages} {101} (\bibinfo {year}
  {2018})},\ \Eprint {http://arxiv.org/abs/1710.00845} {arXiv:1710.00845
  [astro-ph.CO]} \BibitemShut {NoStop}%
\bibitem [{\citenamefont {Riess}\ \emph {et~al.}(2019)\citenamefont {Riess},
  \citenamefont {Casertano}, \citenamefont {Yuan}, \citenamefont {Macri},\ and\
  \citenamefont {Scolnic}}]{Riess:2019cxk}%
  \BibitemOpen
  \bibfield  {author} {\bibinfo {author} {\bibfnamefont {A.~G.}\ \bibnamefont
  {Riess}}, \bibinfo {author} {\bibfnamefont {S.}~\bibnamefont {Casertano}},
  \bibinfo {author} {\bibfnamefont {W.}~\bibnamefont {Yuan}}, \bibinfo {author}
  {\bibfnamefont {L.~M.}\ \bibnamefont {Macri}}, \ and\ \bibinfo {author}
  {\bibfnamefont {D.}~\bibnamefont {Scolnic}},\ }\href {\doibase
  10.3847/1538-4357/ab1422} {\bibfield  {journal} {\bibinfo  {journal}
  {Astrophys. J.}\ }\textbf {\bibinfo {volume} {876}},\ \bibinfo {pages} {85}
  (\bibinfo {year} {2019})},\ \Eprint {http://arxiv.org/abs/1903.07603}
  {arXiv:1903.07603 [astro-ph.CO]} \BibitemShut {NoStop}%
\bibitem [{\citenamefont {{Lesgourgues}}(2011)}]{2011arXiv1104.2932L}%
  \BibitemOpen
  \bibfield  {author} {\bibinfo {author} {\bibfnamefont {J.}~\bibnamefont
  {{Lesgourgues}}},\ }\href@noop {} {\bibfield  {journal} {\bibinfo  {journal}
  {arXiv e-prints}\ ,\ \bibinfo {eid} {arXiv:1104.2932}} (\bibinfo {year}
  {2011})},\ \Eprint {http://arxiv.org/abs/1104.2932} {arXiv:1104.2932
  [astro-ph.IM]} \BibitemShut {NoStop}%
\bibitem [{\citenamefont {{Blas}}\ \emph {et~al.}(2011)\citenamefont {{Blas}},
  \citenamefont {{Lesgourgues}},\ and\ \citenamefont
  {{Tram}}}]{2011JCAP...07..034B}%
  \BibitemOpen
  \bibfield  {author} {\bibinfo {author} {\bibfnamefont {D.}~\bibnamefont
  {{Blas}}}, \bibinfo {author} {\bibfnamefont {J.}~\bibnamefont
  {{Lesgourgues}}}, \ and\ \bibinfo {author} {\bibfnamefont {T.}~\bibnamefont
  {{Tram}}},\ }\href {\doibase 10.1088/1475-7516/2011/07/034} {\bibfield
  {journal} {\bibinfo  {journal} {\jcap}\ }\textbf {\bibinfo {volume} {2011}},\
  \bibinfo {eid} {034} (\bibinfo {year} {2011})},\ \Eprint
  {http://arxiv.org/abs/1104.2933} {arXiv:1104.2933 [astro-ph.CO]} \BibitemShut
  {NoStop}%
\bibitem [{\citenamefont {Torrado}\ and\ \citenamefont
  {Lewis}(2019)}]{torrado_lewis_2019}%
  \BibitemOpen
  \bibfield  {author} {\bibinfo {author} {\bibfnamefont {J.}~\bibnamefont
  {Torrado}}\ and\ \bibinfo {author} {\bibfnamefont {A.}~\bibnamefont
  {Lewis}},\ }\href {https://github.com/CobayaSampler/cobaya} {\enquote
  {\bibinfo {title} {Cobaya},}\ } (\bibinfo {year} {2019})\BibitemShut
  {NoStop}%
\bibitem [{\citenamefont {Lin}\ \emph {et~al.}(2019{\natexlab{b}})\citenamefont
  {Lin}, \citenamefont {Raveri},\ and\ \citenamefont {Hu}}]{Lin:2018nxe}%
  \BibitemOpen
  \bibfield  {author} {\bibinfo {author} {\bibfnamefont {M.-X.}\ \bibnamefont
  {Lin}}, \bibinfo {author} {\bibfnamefont {M.}~\bibnamefont {Raveri}}, \ and\
  \bibinfo {author} {\bibfnamefont {W.}~\bibnamefont {Hu}},\ }\href {\doibase
  10.1103/PhysRevD.99.043514} {\bibfield  {journal} {\bibinfo  {journal} {Phys.
  Rev. D}\ }\textbf {\bibinfo {volume} {99}},\ \bibinfo {pages} {043514}
  (\bibinfo {year} {2019}{\natexlab{b}})},\ \Eprint
  {http://arxiv.org/abs/1810.02333} {arXiv:1810.02333 [astro-ph.CO]}
  \BibitemShut {NoStop}%
\bibitem [{\citenamefont {Mortonson}\ \emph {et~al.}(2009)\citenamefont
  {Mortonson}, \citenamefont {Hu},\ and\ \citenamefont
  {Huterer}}]{Mortonson:2009qq}%
  \BibitemOpen
  \bibfield  {author} {\bibinfo {author} {\bibfnamefont {M.~J.}\ \bibnamefont
  {Mortonson}}, \bibinfo {author} {\bibfnamefont {W.}~\bibnamefont {Hu}}, \
  and\ \bibinfo {author} {\bibfnamefont {D.}~\bibnamefont {Huterer}},\ }\href
  {\doibase 10.1103/PhysRevD.80.067301} {\bibfield  {journal} {\bibinfo
  {journal} {Phys. Rev. D}\ }\textbf {\bibinfo {volume} {80}},\ \bibinfo
  {pages} {067301} (\bibinfo {year} {2009})},\ \Eprint
  {http://arxiv.org/abs/0908.1408} {arXiv:0908.1408 [astro-ph.CO]} \BibitemShut
  {NoStop}%
\bibitem [{\citenamefont {Abbott}\ \emph {et~al.}(2021)\citenamefont {Abbott}
  \emph {et~al.}}]{DES:2021wwk}%
  \BibitemOpen
  \bibfield  {author} {\bibinfo {author} {\bibfnamefont {T.~M.~C.}\
  \bibnamefont {Abbott}} \emph {et~al.} (\bibinfo {collaboration} {DES}),\
  }\href@noop {} {\  (\bibinfo {year} {2021})},\ \Eprint
  {http://arxiv.org/abs/2105.13549} {arXiv:2105.13549 [astro-ph.CO]}
  \BibitemShut {NoStop}%
\bibitem [{\citenamefont {Secco}\ \emph {et~al.}(2022)\citenamefont {Secco},
  \citenamefont {Karwal}, \citenamefont {Hu},\ and\ \citenamefont
  {Krause}}]{Secco:2022kqg}%
  \BibitemOpen
  \bibfield  {author} {\bibinfo {author} {\bibfnamefont {L.~F.}\ \bibnamefont
  {Secco}}, \bibinfo {author} {\bibfnamefont {T.}~\bibnamefont {Karwal}},
  \bibinfo {author} {\bibfnamefont {W.}~\bibnamefont {Hu}}, \ and\ \bibinfo
  {author} {\bibfnamefont {E.}~\bibnamefont {Krause}},\ }\href@noop {} {\
  (\bibinfo {year} {2022})},\ \Eprint {http://arxiv.org/abs/2209.12997}
  {arXiv:2209.12997 [astro-ph.CO]} \BibitemShut {NoStop}%
\bibitem [{\citenamefont {Aiola}\ \emph {et~al.}(2020)\citenamefont {Aiola}
  \emph {et~al.}}]{ACT:2020gnv}%
  \BibitemOpen
  \bibfield  {author} {\bibinfo {author} {\bibfnamefont {S.}~\bibnamefont
  {Aiola}} \emph {et~al.} (\bibinfo {collaboration} {ACT}),\ }\href {\doibase
  10.1088/1475-7516/2020/12/047} {\bibfield  {journal} {\bibinfo  {journal}
  {JCAP}\ }\textbf {\bibinfo {volume} {12}},\ \bibinfo {pages} {047} (\bibinfo
  {year} {2020})},\ \Eprint {http://arxiv.org/abs/2007.07288} {arXiv:2007.07288
  [astro-ph.CO]} \BibitemShut {NoStop}%
\bibitem [{\citenamefont {Choi}\ \emph {et~al.}(2020)\citenamefont {Choi} \emph
  {et~al.}}]{ACT:2020frw}%
  \BibitemOpen
  \bibfield  {author} {\bibinfo {author} {\bibfnamefont {S.~K.}\ \bibnamefont
  {Choi}} \emph {et~al.} (\bibinfo {collaboration} {ACT}),\ }\href {\doibase
  10.1088/1475-7516/2020/12/045} {\bibfield  {journal} {\bibinfo  {journal}
  {JCAP}\ }\textbf {\bibinfo {volume} {12}},\ \bibinfo {pages} {045} (\bibinfo
  {year} {2020})},\ \Eprint {http://arxiv.org/abs/2007.07289} {arXiv:2007.07289
  [astro-ph.CO]} \BibitemShut {NoStop}%
\bibitem [{\citenamefont {Dutcher}\ \emph {et~al.}(2021)\citenamefont {Dutcher}
  \emph {et~al.}}]{SPT-3G:2021eoc}%
  \BibitemOpen
  \bibfield  {author} {\bibinfo {author} {\bibfnamefont {D.}~\bibnamefont
  {Dutcher}} \emph {et~al.} (\bibinfo {collaboration} {SPT-3G}),\ }\href
  {\doibase 10.1103/PhysRevD.104.022003} {\bibfield  {journal} {\bibinfo
  {journal} {Phys. Rev. D}\ }\textbf {\bibinfo {volume} {104}},\ \bibinfo
  {pages} {022003} (\bibinfo {year} {2021})},\ \Eprint
  {http://arxiv.org/abs/2101.01684} {arXiv:2101.01684 [astro-ph.CO]}
  \BibitemShut {NoStop}%
\bibitem [{\citenamefont {Balkenhol}\ \emph {et~al.}(2022)\citenamefont
  {Balkenhol} \emph {et~al.}}]{Balkenhol:2022rvc}%
  \BibitemOpen
  \bibfield  {author} {\bibinfo {author} {\bibfnamefont {L.}~\bibnamefont
  {Balkenhol}} \emph {et~al.},\ }\href@noop {} {\  (\bibinfo {year} {2022})},\
  \Eprint {http://arxiv.org/abs/2212.05642} {arXiv:2212.05642 [astro-ph.CO]}
  \BibitemShut {NoStop}%
\bibitem [{\citenamefont {Lin}\ \emph {et~al.}(2020)\citenamefont {Lin},
  \citenamefont {Hu},\ and\ \citenamefont {Raveri}}]{Lin:2020jcb}%
  \BibitemOpen
  \bibfield  {author} {\bibinfo {author} {\bibfnamefont {M.-X.}\ \bibnamefont
  {Lin}}, \bibinfo {author} {\bibfnamefont {W.}~\bibnamefont {Hu}}, \ and\
  \bibinfo {author} {\bibfnamefont {M.}~\bibnamefont {Raveri}},\ }\href
  {\doibase 10.1103/PhysRevD.102.123523} {\bibfield  {journal} {\bibinfo
  {journal} {Phys. Rev. D}\ }\textbf {\bibinfo {volume} {102}},\ \bibinfo
  {pages} {123523} (\bibinfo {year} {2020})},\ \Eprint
  {http://arxiv.org/abs/2009.08974} {arXiv:2009.08974 [astro-ph.CO]}
  \BibitemShut {NoStop}%
\bibitem [{\citenamefont {Hill}\ \emph {et~al.}(2022)\citenamefont {Hill} \emph
  {et~al.}}]{Hill:2021yec}%
  \BibitemOpen
  \bibfield  {author} {\bibinfo {author} {\bibfnamefont {J.~C.}\ \bibnamefont
  {Hill}} \emph {et~al.},\ }\href {\doibase 10.1103/PhysRevD.105.123536}
  {\bibfield  {journal} {\bibinfo  {journal} {Phys. Rev. D}\ }\textbf {\bibinfo
  {volume} {105}},\ \bibinfo {pages} {123536} (\bibinfo {year} {2022})},\
  \Eprint {http://arxiv.org/abs/2109.04451} {arXiv:2109.04451 [astro-ph.CO]}
  \BibitemShut {NoStop}%
\bibitem [{\citenamefont {Smith}\ \emph {et~al.}(2022)\citenamefont {Smith},
  \citenamefont {Lucca}, \citenamefont {Poulin}, \citenamefont {Abellan},
  \citenamefont {Balkenhol}, \citenamefont {Benabed}, \citenamefont {Galli},\
  and\ \citenamefont {Murgia}}]{Smith:2022hwi}%
  \BibitemOpen
  \bibfield  {author} {\bibinfo {author} {\bibfnamefont {T.~L.}\ \bibnamefont
  {Smith}}, \bibinfo {author} {\bibfnamefont {M.}~\bibnamefont {Lucca}},
  \bibinfo {author} {\bibfnamefont {V.}~\bibnamefont {Poulin}}, \bibinfo
  {author} {\bibfnamefont {G.~F.}\ \bibnamefont {Abellan}}, \bibinfo {author}
  {\bibfnamefont {L.}~\bibnamefont {Balkenhol}}, \bibinfo {author}
  {\bibfnamefont {K.}~\bibnamefont {Benabed}}, \bibinfo {author} {\bibfnamefont
  {S.}~\bibnamefont {Galli}}, \ and\ \bibinfo {author} {\bibfnamefont
  {R.}~\bibnamefont {Murgia}},\ }\href {\doibase 10.1103/PhysRevD.106.043526}
  {\bibfield  {journal} {\bibinfo  {journal} {Phys. Rev. D}\ }\textbf {\bibinfo
  {volume} {106}},\ \bibinfo {pages} {043526} (\bibinfo {year} {2022})},\
  \Eprint {http://arxiv.org/abs/2202.09379} {arXiv:2202.09379 [astro-ph.CO]}
  \BibitemShut {NoStop}%
\bibitem [{\citenamefont {La~Posta}\ \emph {et~al.}(2022)\citenamefont
  {La~Posta}, \citenamefont {Louis}, \citenamefont {Garrido},\ and\
  \citenamefont {Hill}}]{LaPosta:2021pgm}%
  \BibitemOpen
  \bibfield  {author} {\bibinfo {author} {\bibfnamefont {A.}~\bibnamefont
  {La~Posta}}, \bibinfo {author} {\bibfnamefont {T.}~\bibnamefont {Louis}},
  \bibinfo {author} {\bibfnamefont {X.}~\bibnamefont {Garrido}}, \ and\
  \bibinfo {author} {\bibfnamefont {J.~C.}\ \bibnamefont {Hill}},\ }\href
  {\doibase 10.1103/PhysRevD.105.083519} {\bibfield  {journal} {\bibinfo
  {journal} {Phys. Rev. D}\ }\textbf {\bibinfo {volume} {105}},\ \bibinfo
  {pages} {083519} (\bibinfo {year} {2022})},\ \Eprint
  {http://arxiv.org/abs/2112.10754} {arXiv:2112.10754 [astro-ph.CO]}
  \BibitemShut {NoStop}%
\end{thebibliography}%
\end{document}